
\input epsf
\def\slash{/\kern-6pt}
\magnification=1200
\newcount\eqnumber
\eqnumber=1
\def\neweq{\eqno(\the\eqnumber\global\advance\eqnumber by 1)}
\def\eqname#1{\xdef#1{\the\eqnumber}\neweq}
\newcount\refnumber
\refnumber=1
\def\newref{\the\refnumber\global\advance\refnumber by 1}
\def\refname#1{\xdef#1{\the\refnumber}\newref}
\def \psibar{\overline\psi}
\def \fb{\overline f}
\def \hhf{\tilde H}
\def \ktw{\tilde K}

\pageno=1
\rightline{FSU-SCRI-95-101}
\rightline{Oct 1995}
\vskip .5in
\centerline{\bf The Phase Structure of the Schwinger Model on the Lattice}
\centerline{\bf  with Wilson Fermions in the Hartree-Fock Approximation}
\bigskip
\bigskip
\centerline{Ivan Horv\'ath}
\bigskip
\centerline{Supercomputer Computations Research Institute}
\centerline{Florida State University}
\centerline{Tallahassee, FL 32306-4052}
\bigskip
\medskip
\baselineskip=18pt
\centerline {ABSTRACT}
\bigskip
{\narrower The phase diagram of the Schwinger model on the lattice with
one and two degenerate flavours of Wilson fermions is investigated
in the Hartree-Fock approximation. In case of a single
flavour (not directly amenable to numerical simulation), the
calculation indicates the existence of the parity violating phase
at both weak and intermediate-to-strong couplings. In the broken phase,
the Hartree-Fock vacuum sustains a nonzero electric field.
With two flavours, parity is not broken at weak coupling. However,
both parity and flavour become spontaneously broken at the Hartree-Fock
level as the coupling becomes strong.
}
\vfill\eject

\noindent{\bf I. Introduction}
\bigskip
\bigskip

Understanding lattice fermions has been an outstanding issue since
the beginnings of lattice field theory but a satisfactory
insight is still missing. Aside from nontrivial numerical complications
introduced by fermionic degrees of freedom, there are well-known
conceptual obstacles, usually referred to as the ``fermion doubling
problem''.  As revealed by the Nielsen-Ninomiya theorem
[\refname\nogoref], the doubling problem is intimately connected
to chiral symmetry. In fact, this ``No-Go'' theorem has succeeded so far
in preventing us from formulating chiral gauge theory on the lattice,
although the investigations have intensified recently
(see e.g. [\refname\mikeref,\refname\shamref]
and references therein) and possible clues might be at hand.

Except from fundamental importance of chiral symmetry in electroweak
theory, chiral symmetry has long been believed crucial in understanding
the low energy behaviour of strong interactions, described by a
vectorlike theory like QCD. Hinted by the small pion masses,
the basic starting point is that the approximate chiral symmetry crucially
shapes the way the low energy strong interacting world looks.
Pions are regarded as Goldstone bosons coming from the spontaneous
breakdown of this chiral symmetry, and the powerful predictions
of the current algebra follow [\refname\currentalgebraref].

It is desirable to study these interesting issues within
the nonperturbative framework of lattice QCD. Nielsen-Ninomiya theorem
doesn't directly prevent us from ``latticizing'' a vectorlike theory,
even in the chiral limit. However, the existence of the chiral anomaly
complicates the situation considerably. The standard argument is that
since the lattice is a physical regulator, any symmetry of the lattice
action will remain the valid symmetry of the theory at every stage,
including the continuum limit. Consequently, an explicitly chirally
symmetric lattice model can not reproduce the anomaly
structure of the vectorlike
theory. Indeed, the most extensively used versions of lattice QCD,
namely with Wilson fermions and Kogut-Susskind fermions, both
explicitly violate chiral symmetry. Nevertheless, it can be shown in
lattice perturbation theory that the amount of violation is just
right to obtain the correct anomaly in the continuum limit
[\refname\anomref]. This is of course quite comforting.
However, to understand chiral symmetry
on the lattice at finite lattice spacing
(where all the numerical simulations are performed)
means to relate the concepts of chiral symmetry breaking to those
of the lattice system, where chiral symmetry is not present.

Restricting myself now to the case of Wilson fermions, it is generally
believed that there is a line of phase transitions $\kappa_c(g)$
(in the hopping parameter - gauge coupling plane)
running up from the QCD fixed point $\kappa_c(0)$.
On this line, pion-like state becomes
massless and the continuum chiral QCD is believed to be approached
by following this line towards $\kappa_c(0)$. Some time ago
Aoki [\refname\aokiref] set out to answer the questions
raised in the previous paragraph in this context. Namely, since
at $g\ne0$ the masslessness of the pion can't be due to the spontaneous
chiral symmetry breaking, what is it due to? In other words, what is the
nature of the phase transition along $\kappa_c(g)$?

Aoki's answer was that $\kappa_c(g)$ represents the line of
phase transitions at which parity (one-flavour case) or parity
and flavour (multi-flavour case) becomes spontaneously broken.
For one flavour, the pion is identified with the massless particle
driving the parity violating phase transition
($\langle i\psibar\gamma_5\psi\rangle\neq0$). In case of two
flavours, it is $\langle i \psibar\gamma_5 \tau_3\psi\rangle$
acquiring an expectation value. $\pi^0$ is identified with
a massless mode of this phase transition, while the charged pions
are viewed as the Goldstone bosons coming from the breakdown
of flavour. $\langle\psibar\gamma_5 1 \psi \rangle$ is always
assumed to be zero thus giving no reason for $\eta$ to be light,
which could be viewed as a solution of the U(1) problem
on a lattice [\aokiref].

At strong coupling, the supporting evidence for this scenario
is quite convincing [\aokiref]. On the other hand
in the weak coupling regime (relevant for the approach to the
continuum), the situation is far from conclusive. Despite
the fact that some numerical work has been [\refname\aogoref] and
continues to be done [\refname\aoukumref], the very existence of the
parity-flavour violating phase still needs to be examined,
let alone the detailed
picture of symmetry breaking. The existence of the parity-flavour
violating phase has been recently established in the Nambu-Jona-Lasinio
model using large N methods [\refname\abgref] and also in a numerical
simulations with finite number of colours (N=2) [\refname\bivrref].
In that case, the parity-flavour breaking phase in a model with
two flavours seems only to exist at strong and intermediate couplings.

Recently, there has been a line of seemingly unrelated
developments taking place concerning chiral symmetry
following the ideas of Kaplan [\refname\kaplanref]. His approach
amounts to the use of the surface modes of the vectorlike
theory with Wilson fermions in $2d+1$ dimensions as a basis
for the construction of $2d$-dimensional
chiral gauge theory on the lattice.
(For review, see [\refname\jansenref] and references
therein.) Detailed Hamiltonian analysis of the surface
modes of Wilson fermions has been carried out in
[\refname\ourref], where it was also suggested that the
notion of surface modes might be useful for understanding
the conventional Wilson formulation as well. In particular,
one can think of the appearance of the surface modes as an
underlying mechanism generating the parity violating phase.

The nature of the argument is as follows. Consider the Hamiltonian
for one flavour of free Wilson fermions in one spatial
dimension on a finite lattice with open boundaries
$$
H_W = K \sum_j \bigl[ \psibar_{j+1} ( i\gamma_1-r ) \psi_j
                     -\psibar_j ( i\gamma_1+r ) \psi_{j+1} \bigr]
     +M \sum_j\psibar_j \psi_j.        \eqname{\wilham}
$$
Here $K,M,r$ are hopping parameter, mass and the Wilson parameter
respectively. $\psi_j$ is a two component spinor living on site $j$.
If one increases the hopping parameter (or equivalently
decreases the mass) to the supercrticilal values, so that
$$
\Bigl\vert{M \over 2Kr}\Bigr\vert < 1, \neweq
$$
two levels start to behave differently from the rest of the spectrum
and appear bound to the ends of the lattice [\ourref]. As
the size {L} of the system goes to infinity, the energy of these
surface modes tends to zero. On a finite lattice the two
modes mix and acquire the energy $\epsilon\sim e^{-L}$. Consider
now the Dirac vacuum with all the negative energy levels filled.
The last filled level will either be the surface mode on the right
or the one on the left
\footnote*{On a finite lattice I assume that some very small left-right
symmetry breaking term is added to the $H_W$, so that the left and right
modes do not mix.},
thus creating an asymmetric distribution
of particles in the vacuum with an extra particle on one end.
Consequently, after turning on the $U(1)$ gauge field, this
will generate an electric field running through the vacuum. In
one spatial domension such a field can not be canceled by a pair
production and we find that parity is not respected by the vacuum
of this theory.

These ideas were further developed in a recent inspirative review
by Creutz [\mikeref]. He gives a comprehensive
qualitative picture of the phase structure using the surface modes
scenario in both single and multi-flavour case. The argument
is based on the frequently used analogy between nonabelian gauge
theories in four dimensions and electrodynamics in two dimensions
($QED_2$, massive Schwinger model). As is well known, $QED_2$
exhibits some of the most intriguing features ascribed to QCD,
namely confinement, chiral symmetry breaking and the existence
of the $\theta$ parameter. As such it represents a popular
toy model for QCD with the advantage that it can be
analysed semiclassically through bosonisation [\refname\colemanref].
Taking lessons from the continuum and combining them with
the existence of the doublers and surface modes on the lattice,
one naturally arrives to the conclusion that the physics of the
parity violating phase corresponds to the $\theta=\pi$ case
in the continuum. This is based on the considerations of the
previous paragraph and the fact that in $QED_2$ the $\theta$
parameter has a direct physical meaning as the background
electric field. The phase diagrams of Ref.~[\mikeref]
represent the expected positions of $\theta=\pi$ transitions
on the lattice that should be applicable at weak coupling.

The purpose of this paper is to investigate the issues of
a parity violating phase in lattice $QED_2$ with Wilson
fermions in a direct lattice calculation. Compared to the
wealth of exact and approximate information
accumulated over the years on the continuum Schwinger model
\footnote*{The recent activity concentrated mostly on
the multi-flavour case. See e.g. [\refname\schconref].},
the knowledge we have on the lattice
is rather modest. First of all, there are no exact solutions
within any of the formulations where doublers are removed.
The model was mostly investigated with Kogut-Susskind fermions,
testing various numerical methods by comparing the lattice results
to the exactly known continuum quantities in the massless case
(see e.g. [\refname\ksschw]). However, very little
is known about the theory in the Wilson fermion formulation.
The situation is particularly interesting for a single
flavour, because in that case the direct numerical
simulation is not possible. This is due to the fact that
for certain gauge configurations the fermionic determinant
is not positive and therefore one does not have a probabilistic
weight for the purposes of Monte Carlo simulation.
In a recent work [\refname\galaref], Gausterer and Lang
studied the Lee Yang zeros of the partition function
on small lattices analytically (at infinite coupling)
and numerically (at intermediate couplings).
The system was also studied in Ref.~[\refname\azcoitiref].
In case of two flavours, the direct numerical analysis is
possible, but to my knowledge, the systematic study of phase
structure has not been carried out.

In what follows, I will study the phase diagram of Hamiltonian
$QED_2$ on the lattice with one and two degenerate flavours of Wilson
fermions in the Hartree-Fock (H-F) approximation. I will work in the
axial gauge, where the gauge degrees of freedom are easily
eliminated in favour of fermionic fields, thus providing
a convenient setup for the use of the H-F approximation to
the vacuum of the theory. The H-F ground state, or the independent
fermion ground state, is a state one might call
(by definition) the mean field ground state for the theory
of interacting, particle number conserving (charge
conserving) fermions. It has a nice
variational interpretation and can be regarded as a first
term in a series of systematic variational
improvements. While belonging to the standard set of techniques
used in many-body theory, H-F methods are rarely invoked in the
lattice gauge context.
In the case under consideration, however, it can give us
valuable hints about the phase structure of the theory.

Investigating the phase diagram, I adopt the point of
view, taken in Ref.~[\mikeref], that it is quite natural to
introduce the axial mass term (the ``$M_5$''-term).
More specifically, I will consider the generalized
version of Hamiltonian (\wilham), given by
$$
H_{W_5} = H_W + M_5 i\psibar_j\gamma_5\psi_j
\equiv H_W + H_5.  \eqname{\wil5ham}
$$
Indeed, one obtains such a term from the conventional mass term
by a chiral rotation. The existence of a $\theta$ parameter
in the continuum theory can be thought of as being due
to the fact that because of regularization, the theory with
such a rotated mass term is actually not equivalent to the
original one and is thus anomalous. In this way, introducing
$M_5$ essentially means trading $\theta$
(being an independent parameter of the theory)
in favour of this new mass term. Consequently, I will consider
the lattice theory in the space of three bare parameters,
$M,M_5$ and $g$ (gauge coupling), in contrast to three
conventional continuum parameters $m,\theta$ and $e$.

In Sec.~II, starting from the continuum theory in axial gauge,
I will formulate the lattice model with one flavour and discuss
its discrete symmetries. The H-F approximation is then described
in Sec.~III. I do this in some detail, stressing the variational
character of the method and the fact that all the discrete
symmetries (including parity) are preserved by the approximation.
The numerical solutions of the H-F equations on finite lattices
and their implications on the H-F phase diagram are discussed
in Sec.~IV. I try to make all the observed qualitative features
of the vacuum plausible by tracing them to the elementary picture
of interacting particles in filled H-F levels. The model with
two flavours is then formulated and analysed in Secs.~V and VI.
Summary, together with some generalizations and speculations,
is given in Sec.~VII. Finally, the numerical procedure used
to solve the H-F equations is described in Appendix.

\bigskip
\bigskip
\noindent {\bf II. The Model with One Flavour}
\bigskip
\bigskip

The massive Schwinger model with one flavour of fermions is defined
by the Lagrangian
$$
{\cal L} = -{1\over 4}F_{\mu\nu}F^{\mu\nu} +
  \psibar\bigl[\gamma^\mu(i\partial_\mu  - eA_\mu) - m \bigr]\psi \,,
\neweq
$$
where $F_{\mu\nu}=\partial_\mu A_\nu - \partial_\nu A_\mu$.
$\psi$ is the two-component spinor field, $A_\mu$ is the gauge
field and $m,e$ the mass and dimensionful coupling constant respectively.
I will consider the theory in axial gauge $(A_1=0)$.

The equation of motion for $A_0$ is the
equation of constraint (Gauss' law)
$$
E'(x)\equiv -A_0''(x) = e\psi^\dagger(x)\psi(x) \equiv e\rho(x) \,,
\eqname{\gauss}
$$
where $E(x)\equiv -A_0'(x)$ is the electric field and $\rho$ the
charge density. It can be solved by
$$
A_0(x) = - {1\over 2} \int dx' \mid x-x'\mid e \rho(x') + Bx + C \,.
\neweq
$$
The integration constant $C$ is physically irrelevant and will be put
to zero. The constant $B$ represents a uniform background field
which can be also put to zero for my current purposes.
Then the Hamiltonian of the theory for the zero total charge
takes the form
$$
H= \int dx \Bigl[ \psibar(i\gamma_1\partial_1 +m )\psi
   - {e^2 \over 4} \int dx' \rho(x)\mid x-x' \mid \rho(x') \Bigr] \,.
\neweq
$$

Upon quantization, $\psi$ and $H$ become operators in the
corresponding Hilbert space, with field operators subject to the canonical
anticommutation relations. In the standard treatment, the local
charge density operator is replaced by its normal ordered version
$\colon\psi^\dagger\psi\colon$, with the normal ordering usually performed
with respect to the filled Dirac sea (no particles, no antiparticles).
This effectively amounts to the compensation of the infinite charge,
generated by the sea. Since I will not introduce the antiparticle
operators it is more convenient for my later purposes to define the
charge density in such a way that the compensation is explicit,
namely
$$
\rho(x) \equiv \psi^\dagger(x)\psi(x) - 1 \,.   \eqname{\rodef}
$$

To formulate this theory on a lattice is now straightforward.
It will be defined by the Hamiltonian
$$
H = H_{W_5} - {g^2 \over 4} \sum_{n,m}\rho_n\mid n-m \mid \rho_m
\equiv H_{W_5} + H_I \,,
\eqname{\gsham}
$$
where $H_{W_5}$ is the free part with $H_5$ defined in (\wil5ham) and
$$
\rho_n = \psi^\dagger_n\psi_n - 1 \,.
\eqname{\chargedef}
$$
The indices $n,m$ label the lattice sites runing from $1$ to $L$
and the lattice spacing has been set to unity. Fermionic
variables are subject to open boundary conditions
and satisfy the canonical anticommutation relations
$$
\lbrace \psi_n^{\sigma},\psi_m^{\dagger\tau}\rbrace
 = \delta_{nm}\delta_{\sigma\tau} \, , \neweq
$$
with $\sigma,\tau$ being the spinor indices. The electric field
in this formulation is a derived quantity, defined through the
the lattice analog of Gauss' law (\gauss) by
$$
E_j \,=\, {g\over 2}\, \Bigl[\, \sum_{l=1}^j\rho_l\,
                            - \sum_{l=j+1}^L\rho_l\,\Bigr] \,.
\eqname{\elfield}
$$
Here $E_j$ is the operator of electric field on link $(j,j+1)$.

It is worth emphasizing at this point that similarly to the continuum
case, the lattice interacting theory posesses exact discrete
symmetries. Choosing the representation of $\gamma$ matrices as
$$
\gamma^0=\pmatrix{0&1\cr
                  1&0\cr} \qquad \quad
\gamma^1=\pmatrix{0&1\cr
                 -1&0\cr} \qquad \quad
\gamma_5=\gamma^0\gamma^1 \,,
\neweq
$$
the operations of C,P,T in continuum (left column) and on the lattice
(right column) are defined by the following transformation properties of the
field operators
$$
\eqalign{P\psi (x) P^{-1}&=\gamma_0\psi (-x)\cr
         C\psi (x) C^{-1}&=\gamma_1\psibar^T (x)\cr
         T\psi (x) T^{-1}&=\gamma_0\psi (x)\cr}
\qquad \qquad \qquad
\eqalign{P\psi_j P^{-1}&=\gamma_0\psi_{L+1-j}\cr
         C\psi_j C^{-1}&=\gamma_1\psibar^T_j\cr
         T\psi_j T^{-1}&=\gamma_0\psi_j\,.\cr}
\eqname{\transforms}
$$
Here $C,P$ are unitary and $T$ antiunitary operators. The above
transformations indeed leave the corresponding continuum and lattice
Hamiltonians without $H_5$ ($M_5=0$) unchanged.

Note that $H_5$ is odd under $P,C$ and even under $T$. Therefore $CP$ and
$T$ are the exact symmetries of the theory at any $M_5$. This is
different from the situation in four dimensions, where $H_5$ is invariant
under $C$ and changes sign under $P,T$, which in turn implies that both
$PC$ and $T$ are explicitly broken by nonzero $M_5$.

\bigskip
\bigskip
\noindent {\bf III. Hartree-Fock Approximation}
\bigskip
\bigskip

In solid state physics, the H-F approximation is frequently referred to
as the ``independent electron approximation'' and this probably
captures its essence best. Indeed, the main idea is to approximate
a given state of the fermionic many-body system by a Slater
determinant of some set of one-particle states. This is usually applied
to approximate the unknown vacuum of the theory, which is also
my main interest here. In that case the method boils down to
finding a set of one-particle states with the Slater determinant of
minimal energy.

Let me therefore start to investigate the vacuum of the one-flavour
lattice Schwinger model by considering an arbitrary (but fixed)
complete orthonormal set of one-particle fermionic states
on a lattice of $L$ sites
$$
S \equiv \{\,\phi^\alpha\, \mid\, \alpha = 1,2,\ldots,2L\,\}. \neweq
$$
Every state $\phi^\alpha$ is a collection of two-component spinors,
residing on site $n$
$$
\phi^\alpha\equiv \{\,\phi_n^\alpha\,\mid\, n = 1,2,\ldots,L\,\}
\qquad \qquad
\phi^\alpha_n \equiv \pmatrix{\phi^{\alpha,1}_n\cr \phi^{\alpha,2}_n\cr}.
\neweq
$$
One can build the fermionic many-body Fock space out of these states
in a standard way and define the complete set of fermionic annihilation
operators by
$$
\psi_n = \sum_\alpha a_\alpha \phi^\alpha_n \quad .
\eqname{\fedec}
$$

With definition (\chargedef) of the local charge operator, restriction
to the charge zero sector translates into the requirement of half-filling.
In other words, only states with $L$ particles are allowed. An arbitrary
state $\mid\psi\rangle$ from this subspace can be written in the form
$$
\mid\psi\rangle \,=\, c^0\mid 0f, 0\fb\,\rangle
                \,+\, \sum_r c^1_r\mid 1f, 1\fb\,\rangle_r
                \,+\, \sum_r c^2_r\mid 2f, 2\fb\,\rangle_r
                \,+\, \ldots \,+\, c^L\mid Lf, L\fb\,\rangle.
\eqname{\stdec}
$$
Here $\mid 0f, 0\fb\,\rangle$, the ``sea'', is the state with first $L$
levels filled
$$
\mid 0f, 0\fb\,\rangle = a^\dagger_1 a^\dagger_2 \ldots a^\dagger_L
                         \mid 0\,\rangle.
\eqname{\slater}
$$
$\mid nf, n\fb\,\rangle$ generically represents states with $n$ fermions
removed from the sea and put into $n$ empty levels ($n$ fermion - $n$
antifermion states). Index $r$ enumerates these basis states at fixed $n$.
Obviously, the dimension of the charge zero sector is $2L\choose L$.

Apart from charge conservation, no other symmetries are assumed to be
respected by the vacuum. One usually restricts the space of states
further by going to subspace of zero momentum. However, momentum is not well
defined on a lattice with open boundaries. Specification
of $2L\choose L$ complex coefficients $c^k_r$ in the above
decomposition therefore constitutes an {\it exact} representation
of the vacuum. Requirement of minimal energy defines a variational
problem for determination of these coefficients. However, on a reasonably
sized lattices the number of variables in the problem becomes too
huge to be manageable.

The variational philosophy behind the H-F approach is to regard the set $S$
as a collection of variational parameters instead. In particular, the
method aims at adjusting the one-particle basis in such a way that while
retaining only the first term in decomposition (\stdec), the lowest energy
state is achieved. Note that in this way, the variational problem
involving $2{2L\choose L}$ real variables is replaced by one involving
$3L^2$ real variables
\footnote{*}{Counting here includes the fact that only the filled levels
represent the true variational variables since only they contribute to the
total energy. One also has to take into account the orthonormality
constraints.}. Moreover, fixing $S$ by the Hartree-Fock
prescription transforms the decomposition (\stdec) into a well
defined variational improvement scheme. Indeed, employing the H-F basis
one expects the states $\mid nf,n\fb\,\rangle$ to play increasingly
less important roles in the true vacuum with increasing
$n$. This is expected to be true regardless of the value of the gauge
coupling.

Having defined the H-F approximation it is now a straightforward matter
to transform the problem into the familiar manageable form. Using the
field decomposition (\fedec)
one can rewrite the free part of the Hamiltonian (\gsham) as
$$
H_{W_5} \,\equiv\, \sum_{nm}\psi^\dagger_n K_{nm} \psi_m \,=\,
\sum_{\alpha\beta} a^\dagger_\alpha {\cal K}^{\alpha\beta} a_\beta \quad,
\neweq
$$
where
$$
\eqalign{ {\cal K}^{\alpha\beta} &=
          \sum_{nm} \phi^{\alpha\dagger}_n K_{nm}\phi^\beta_m\cr
          K_{nm} &= K\bigl[ \delta_{n,m+1}\gamma_0 (i\gamma_1-r)
                           -\delta_{n+1,m}\gamma_0 (i\gamma_1+r)\bigr]
                  + \delta_{nm}\bigl[M\gamma_0 + iM_5\gamma_1\bigr].\cr}
\eqname{\knmdef}
$$
The interaction term becomes a little more complicated and has three
parts
$$
H_I \,=\,{\cal E}_I
    \,+\,{g^2\over 2}\sum_{\alpha\beta} a^\dagger_\alpha
                          {\cal N}^{\alpha\beta}a_\beta
    \,+\,{g^2\over 4}\sum_{\alpha\beta\gamma\delta}
                          a^\dagger_\alpha a^\dagger_\beta
    {\cal M}^{\alpha\beta\gamma\delta} a_\gamma a_\delta \quad,
\neweq
$$
where
$$
{\cal N}^{\alpha\beta} = \sum_{nm}\phi^{\alpha\dagger}_n
                         \mid n-m \mid\phi^\beta_m
\qquad\quad
{\cal M}^{\alpha\beta\gamma\delta} = \sum_{nm}
         \phi^{\alpha\dagger}_n \phi^\gamma_n
         \mid n-m \mid \phi^{\beta\dagger}_m\phi^\delta_m
\neweq
$$
and ${\cal E}_I$ is an unimportant constant
${\cal E}_I=-{g^2\over 4}\sum_{nm}\mid n-m \mid$.
Note that loosely speaking, the constant term corresponds to the
self-interaction of the Dirac sea-compensating charge,
the quadratic term arises due to the interaction of this charge
with the system and the quartic term represents the
interactions of the system itself.

The mean energy in the Slater determinant (\slater) is a function
of the set $S$ and is given by
$$
{\cal E}(S) \,=\, {\cal E}_I \,+\,
                 \sum_{\alpha=1}^L \bigl[ {\cal K}^{\alpha\alpha}
               +{g^2\over 2}{\cal N}^{\alpha\alpha} \bigr]
             \,+\, {g^2\over 4}\sum_{\alpha\beta=1}^L
                  \bigl[ {\cal M}^{\alpha\beta\beta\alpha}
              - {\cal M}^{\alpha\beta\alpha\beta}\bigr].
\eqname{\energy}
$$
The Hartree-Fock set of states $S^{HF}$ is now determined by minimizing
${\cal E}(S)$ with variables $\phi^\alpha_n$ subject to the orthonormality
constraints. Standard manipulations then reveal that this variational
problem can be solved by subjecting the one-particle wavefunctions
to the H-F equations of the form
$$
\sum_m \hhf_{nm}\phi^\alpha_m \,=\, \epsilon^\alpha\phi^\alpha_n \quad ,
\eqname{\hafoeq}
$$
with
$$
\hhf_{nm} \,=\, K_{nm} \,+\,{g^2\over 2}\Bigl[\, V^D_{nm} + V^E_{nm} \,\Bigr]
\neweq
$$
and
$$
V^D_{nm} =  \delta_{nm}\sum_j\mid n-j\mid
            (1-\sum_{\beta =1}^L \phi^{\beta\dagger}_j\phi^\beta_j)\qquad
V^E_{nm} = \sum_{\beta =1}^L\phi^\beta_n\mid n-m\mid
           \phi^{\beta\dagger}_m \, .
\neweq
$$
The ``twidle'' in $\hhf_{nm}$ serves to denote the fact that these are
not the one-particle matrix elements of Hamiltonian (\gsham). As expected,
the H-F equations take the form of a one-particle Hamiltonian eigenstate
problem with the complication that the Hamiltonian matrix depends on the
eigenstates themselves. Thus the equations have to be solved
self-consistently. As usually, the self-consistent potential has direct
and exchange parts.

Due to the self-consistent feature of the above H-F equations, finding an
exact solution is a nontrivial task and I haven't succeeded in doing that.
On the other hand, there is a simple way to attempt to solve these
equations on finite lattices numerically, namely by iteration.
Straightforward application of the iterative procedure however
doesn't converge to the self-consistent solution. The nature of the
problems is similar to those described in Ref.~[\refname\negeleref]
in the context of continuum $QCD_2$ in th large $N$ limit.
I discuss these technical issues in Appendix. Using a modified
approach, numerical solutions can be iteratively
found in wide range of coupling constants.

Let me close this section with a few remarks concerning symmetry within
the H-F
approximation. It is usually helpful and desirable that the approximation
scheme retains as much symmetry of the approximated system as possible.
Especially if the main purpose of the investigation is spontaneous
symmetry breaking. In particular, as discussed in the previous section,
if $M_5=0$, the Hamiltonian (\gsham) is invariant under parity.
Is the parity invariance present in the approximation? In what sense?

The underlying dynamics driving the H-F approximation is entirely
embodied in the H-F equations. Therefore, the symmetries of these
equations should also determine the symmetries of the approximation.
It follows from transformation properties (\transforms) that under
the operation of parity the one-particle wave function $\chi_n$
transforms into $\gamma_0\chi_{L+1-n}$. Since the H-F Hamiltonian
implicitly depends on its eigenstates, the parity operation in this
case has to involve the whole set $S$. In fact, it can be checked
quite easily that the operation
$$
S^{HF}=\{\phi^\alpha_n\} \,\longrightarrow\,
PS^{HF}=\{\gamma_0\phi^\alpha_{-n}\}\,,
\qquad\qquad
\epsilon^\alpha \,\longrightarrow\, \epsilon^\alpha
\neweq
$$
is a symmetry of the H-F equations. In other words, if $S^{HF}$ is a
self-consistent set solving (\hafoeq), then $PS^{HF}$ is also
a self-consistent set with corresponding one-particle energies equal.

Let me also mention that not just parity but all the discrete symmetries
discussed in the previous section are preserved by the H-F approximation
in the above sense. Of course, performing the symmetry operation
on the H-F vacuum can lead to a Slater determinant involving different
self-consistent set, thus opening the possibility of spontaneous
symmetry breaking.

\bigskip
\bigskip
\noindent {\bf IV. Numerical Analysis (One Flavour)}
\bigskip
\bigskip

In this section, I will discuss the results of the H-F analysis for the
model with one flavour. I will concentrate on the phase diagram in the
$M-g^2$ plane. To observe the parity violating effects on a finite
lattice, I fix $M_5$ to a very small value $(M_5=10^{-3})$ throughout
this section. The values of hopping parameter $(K=1)$ and Wilson
parameter $(r=0.5)$ are set in such a way that the critical value
of $M$ at zero coupling is $M_{c}(0)=1$. All quantities are given
in the lattice units. Moreover, the electric field is allways measured
in units of $g$. In particular, this expectation is calculated
using formula (\elfield) with operators of charge density
replaced by their expectation values in the H-F vacuum,
$$
\langle \rho_n \rangle^{HF} = \sum_{\alpha=1}^L
                     \phi^{\alpha\dagger}_n\phi^\alpha_n \,-\, 1\,,
\neweq
$$
and the factor of $g$ removed.

To assess the accuracy of the self-consistent numerical solution and
to provide the criterion for terminating the iterative procedure,
I have computed
$$
\delta^{(k)} \,=\, \sum_{\alpha,n}
                   \mid\, \phi^{\alpha (k+1)}_n - \phi^{\alpha (k)}_n\,\mid
\eqname{\error}
$$
at each iterative step $k$. Here $\phi^{\alpha (k)}_n \in S^{(k)}$,
the one-particle set after $k$ iterations (see Appendix).
Obviously, $\delta = 0$ only for the self-consistent set.
In all cases discussed in this section $S^{(k)}$
has been accepted as a numerical solution only if $\delta^{(k)}< 10^{-2}$.
In most cases however, this number has been much smaller (up to four
orders of magnitude at weak couplings). With the above bound,
the physical characteristics of the H-F Slater determinant
(such as energy), became essentially insensitive to further decrease
of $\delta$. To achieve this accuracy on the lattices
I have studied $(L=32,40,48)$ took typically a few tens of iterations
at weak couplings $(g^2\le 0.5)$ and a couple of hundreds at intermediate
and strong couplings $(g^2> 0.5)$. Working in the vicinity
of the phase transition typically added roughly one order of magnitude
to the number of iterations. In the region of couplings studied here
$(g^2\le 3)$, the self-consistent solution has always been
straightforwardly found with free wave functions at given $M$ used
as a starting point for the iteration.

The representative example of the most relevant finding in this study
is displayed in Fig.1. The vacuum expectation value of the electric
field in the middle of the $32$-site lattice is plotted as a function
of the fermion mass at weak coupling $(g^2=0.1)$.
Note that for large values of $M$, the electric field tends to zero
as one would expect in the parity-invariant theory.
However, at small fermion masses the field acquires an expectation
value and the two regions are separated by a rapid transition.
This suggests the existence of a parity violating phase transition
and confirms the qualitative picture presented in Ref.~[\ourref]
at the Hartree-Fock level.

The spatial dependence of the electric field across the lattice
is plotted in Fig.2a for typical cases in the broken and symmetric
phases. Note that in the broken phase, the field nicely settles
to a uniform bulk value essentially across the whole lattice.
In the symmetric example, the field is almost zero everywhere.
It is quite interesting to see the spatial distribution of
Hartree-Fock levels in these two situations.
This is shown in Fig.2b where I plot the energy of these
levels against the mean position of particles in them.
In the symmetric case, all the particles reside on average
in the middle of the lattice and the left-right symmetry
is preserved up to small explicit violations caused by the
presence of the small $M_5$-term. The filled levels produce a uniform
charge distribution neutralized by the compensating charge.
This is to be compared to the situation in the broken phase,
where the left-right symmetry is completely lost. Indeed, it is
energetically favourable for the levels to spread out asymmetrically.
Filling the sea generates the surface charge and an electric
field.

Similar behaviour is observed also at higher values of the
gauge coupling. The resulting positions of phase transitions
observed on the lattice with $48$ sites are plotted in
Fig.3. The transition points here are determined simply
as the locations of the rapid rise of the vacuum expectation
value of the electric field at fixed coupling. In particular,
the phase transition is assumed to happen at the fermion
mass $M_{c}(g^2)$, where this expectation value rises above
$10^{-2}$, i.e. above the value one order of magnitude larger
than the size of the parity violating $M_5$-term. By comparing
to the results on smaller lattices ($L=32,40$) I expect the
critical masses at nonzero couplings to be increased
by a few parts per hundred in the infinite volume limit.

For the model in the standard Euclidean formulation, Gausterer
and Lang [\galaref] concluded the existence of a phase transition
at infinite coupling. After appropriate rescaling the
parameters of their model, the quoted position of
this transition is $M_{c}(\infty)\simeq 0.32$. Although
I don't know of any apriori reason why the phase transition
should occur at the same place in both formulations,
it is interesting to observe that their result is an
acceptable asymptotic value at strong coupling here too.

Similarly to the electric field, the simplest local fermionic
parity-odd operator, namely axial charge density
$i\psibar\gamma_5\psi$, also acquires an expectation
value at the parity violating phase transition.
This is illustrated in Fig.4a where I plot both the electric
field and the axial charge density as a function of
fermion mass at $g^2=1.0$.
Both operators appear to acquire an expectation value
simultaneously as expected. Typical spatial dependence of
the axial charge density in the broken and symmetric phases
is plotted in Fig.4b, showing the bulk nature of the order
parameter.

The relative size of the electric field and the axial charge
density in the broken phase varies with gauge coupling.
This is demonstrated in Fig.5 where I plot these expectation
values at fixed fermion mass. Note that while the electric field
starts up finite at weak coupling and decreases monotonically
at intermediate and strong couplings, axial charge density
behaves in a complementary way. It approaches zero
with vanishing coupling and rises as the coupling increases.

The above behaviour of electric field is simply a
manifestation of charge shielding, an effect well known
to be present in the continuum theory as well. Indeed,
consider first the free theory. In that case it is just
the filled surface mode that is responsible for the
parity-breaking effects. That's why the electric
field approaches value 0.5 $(\theta=\pi)$ when coupling
tends to zero. However, once the gauge coupling is turned on,
the rest of the levels spread out (see Fig.2b) and
the accumulation of surface charge is a result of the collective
action of all self-consistently interacting particles in
filled states. The net effect of this phenomenon is a
screening of the surface charge. As the value of the gauge
coupling increases, while remaining in the broken phase, one
expects the levels to spread out and screen even more since the
system wants to reduce the positive attraction energy
of the surface charges. For example, at $g^2=1.5$, the
spatial distribution of H-F levels is shown in Fig.6.
At strictly strong coupling, when the interaction
term absolutely dominates, every particle in the sea will live
bound to just one site of the lattice, thus eliminating the
surface charge completely. Therefore, the field is expected
to vanish in this limit in the H-F approximation.

In the light of the above considerations, behaviour of the
axial charge density in Fig.5 becomes also quite natural. Indeed,
since at zero coupling the parity violation is all concentrated
on the ends, it will not be reflected in the expectation value of
the local operator inside the system. Consequently, one expects
the bulk axial density to vanish. At strong coupling however,
parity violation is equally contributed by all the filled
levels and the axial density acquires an expectation value.

An interesting feature already present in the examples
of Fig.2, but quite striking in Fig.6, is that because
of the interaction energy, it is not necessarily the
lowest one-particle states that are filled
to form the H-F vacuum. Indeed, in Fig.6
almost half of the filled levels (denoted by diamonds)
are those with positive one-particle energies. This is
discussed in more detail in the Appendix.

Finally, let me close this section by discussing the order
of the parity violating phase transition in the H-F
approximation. It is well known that the mean field-like
approaches are frequently misleading about the order
of the phase transition and the critical exponents.
Therefore, while I assume that it is plausible for
H-F approximation to recognize the transition,
the information it gives about the order should be
taken with some care.

I have calculated the connected correlation functions
for electric field and axial charge density in the
H-F vacua. For the case of electric field, these functions
have a very nice exponential decay in all cases I have studied
and the corresponding correlation lengths could
be reliably determined. Typical behaviour of the inverse
correlation length (mass gap) across the phase transition
is shown in Fig.6a. In the symmetric phase, the correlators
of the axial charge behave in the same way. However, in the
broken phase they show some differences as can
be observed from Fig.6a. For weak couplings, the determination
of the correlation length from these axial charge correlators
in the broken phase was less accurate than from the electric ones.
Using the electric field correlation functions, Fig.6b shows
the mass gap along the line of phase transitions on a lattice
of $48$ sites. These results are reasonably finite-size
stable (more so at stronger couplings) and I exclude the
possibility of mass gap reducing to zero in the infinite volume limit.
I conclude that in the H-F approximation the phase transition
is of first order at finite $g$, approaching a second order
endpoint at zero coupling.

\bigskip
\bigskip
\noindent {\bf V. Two Flavours}
\bigskip
\bigskip

Turning now to the case of two degenerate fermion flavours,
I will consider the lattice Hamiltonian
$$
H =  \sum_f\sum_{nm}\psi^{f\dagger}_n K_{nm} \psi^f_m  -
     {g^2 \over 4} \sum_{fh}\sum_{n,m}
     \rho^f_n\mid n-m \mid \rho^h_m \,.
\eqname{\twoflham1}
$$
Here $f,h$ are flavour indices assuming two values, $K_{nm}$ is
defined in (\knmdef) and
$$
\rho^f_n = \psi^{f\dagger}_n\psi^f_n - 1 \,.
\neweq
$$
The fermionic operators $\psi^f_n$ are subject to the canonical
anticommutation relations.

Note that with flavours being degenerate, the parameter space of
this two-flavour theory is the same as for a single flavour,
namely $M,M_5,g$. Also, similarly to the one-flavour case,
the above lattice model retains all the discrete symmetries of
the corresponding continuum theory. In addition, the two-flavour model
is invariant under unitary transformations in flavour space.
All of these symmetries will be preserved by the Hartree-Fock
approximation in the sense discussed in Sec.~III.

Inclusion of the flavour index does not require any conceptual changes
in the application of the Hartree-Fock procedure. On the technical
side, it is easiest to skip the explicit use of flavour notation and
assemble the two fermionic operators on site $n$ into a $4-$component
column $\psi_n \equiv {\psi^1_n\choose\psi^2_n}$. The Hamiltonian then
takes the form
$$
H =  \sum_{nm}\psi^\dagger_n \ktw_{nm} \psi_m  -
     {g^2 \over 4} \sum_{n,m}
     \rho_n \mid n-m \mid \rho_m \,,
\eqname{\twoflham2}
$$
with
$$
\ktw_{nm} = \pmatrix{K_{nm}&0\cr 0&K_{nm}}
\neweq
$$
and
$$
\rho_n = \psi^\dagger_n\psi_n - 2 \,.
\neweq
$$
Note that to compensate for the charge of the Dirac sea now requires
two units of charge per site.

Using the above notation and the decomposition
$$
\psi_n = \sum_{\alpha=1}^{4L} a_\alpha\phi_n^\alpha \, ,
\neweq
$$
the derivation of the H-F equations is a line by line repetition
of the procedure for the one flavour case up to the factors of two
coming from the doubling of the compensating charge. Indeed, the H-F
equations take the form
$$
\sum_m \hhf_{nm}\phi^\alpha_m \,=\, \epsilon^\alpha\phi^\alpha_n \quad ,
\eqname{\hafoeq}
$$
with
$$
\hhf_{nm} \,=\, K_{nm} \,+\,{g^2\over 2}\Bigl[\, V^D_{nm} + V^E_{nm} \,\Bigr]
\neweq
$$
and
$$
V^D_{nm} =  \delta_{nm}\sum_j\mid n-j\mid
            (2-\sum_{\beta =1}^{2L} \phi^{\beta\dagger}_j\phi^\beta_j)\qquad
V^E_{nm} = \sum_{\beta =1}^{2L}\phi^\beta_n\mid n-m\mid
           \phi^{\beta\dagger}_m \, .
\neweq
$$
Note that these equations are formally almost identical to those for one
flavour. The crucial difference however is that $\phi_n^\alpha$ is
now a $4-$component object and $\hhf$ a $4L\times 4L$ complex matrix.

\bigskip
\bigskip
\noindent {\bf VI. Numerical Analysis (Two Flavours)}
\bigskip
\bigskip

I will start with the discussion of the phase structure in the $M-M_5$ plane
at fixed gauge coupling. Before turning to the results of the H-F analysis,
let me first briefly explain what one would expect to be happening here at
weak coupling based on the surface mode picture. To do that, I will need
to borrow the of ideas of Ref.~[\mikeref], and to make this paper
reasonably self-contained,
to review briefly the part that is relevant here.
Thinking first in the continuum context, consider the standard mass term
$m\psibar\psi$ and its transformation under the chiral rotation
$\psi\rightarrow e^{i{\theta/2\gamma_5}}\psi$. We have
$$
m\psibar\psi \longrightarrow m\cos(\theta)\psibar\psi
                           + m\sin(\theta)i\psibar\gamma_5\psi \,.
\neweq
$$
Therefore, the chiral rotation by angle $\theta$ corresponds to the
rotation of the vector $(m,0)$ in the $m-m_5$ plane around the origin
(``chiral point'') by the same angle $\theta$. While naively expecting
that the physics should be the same after the above change of variables,
this is actually not the case because of the chiral anomaly. What we are
actually getting is a physics with different ``gauge'' $\theta$-parameters,
i.e. with different background electric
field, realized for example through the existence of the surface charges.
With this identification, the above transformation prescription gives the
approximate relation (the renormalization effects, for example, will shift
the chiral point to negative $m$) between the theory considered in the
parameter spaces $(e,m,\theta)$ and $(e,m,m_5)$.

On a lattice with Wilson fermions, the situation is a little more
complicated,
because except from $(M_{c},0)$, there is another chiral point in the
$M-M_5$ plane, namely $(-M_{c},0)$, where the doubler goes massless.
The conjecture then
is that here the total $\theta$-parameter gets two contributions, each
being the angle with respect to the two chiral points, with doubler's
contribution taken with the reversed sign.
These angles are sketched in Fig.8a. $N_f$ degenerate flavours
will contribute equally to the total value of $\theta$
and consequently, one expects the following approximate relation to hold
at weak coupling
$$
\theta = N_f(\theta_p - \theta_d) \, .
\eqname{\thedef}
$$
Here $\theta_p,\theta_d$ are the contributions of the particle and the doubler
respectively. In other words, on a lattice with open boundaries, the system
is expected to generate surface charges in such a way, that the resulting
electric field will approximately correspond
to $\theta$ given by the above relation.

If this qualitative picture is correct, there should be phase transitions
occuring in the $M-M_5$ plane at the positions
where $\theta$ reaches $\pi$. Indeed, for $\theta>\pi$, it will
be energetically favourable to create a fermion-antifermion pair
thus reducing the magnitude of the electric field
and switching its sign. Consequently, $\theta$ should jump from $\pi$ to
$-\pi$ across these phase transitions. The condition $\theta=\pi$ defines
a line in $M-M_5$ plane, but its qualitative behaviour strongly depends
on the number of flavours. With single flavour, the only solution
is a straight line, connecting $M_c$ and $-M_c$. This can
be understood already from the point of view of the ``naive'' surface
mode picture as I presented in the Introduction.
Indeed, switching the sign of a small parity violating ``$M_5$''-term
causes the two surface modes to exchange the ends of the lattice,
thus switching the sign of the surface charges and the electric field.
For two flavours, the condition $\theta=\pi$ defines a circle with
centre at the origin and radius $M_{c}$. Including a small flavour
breaking in both $M$ and $M_5$ to visualise the chiral endpoints, the phase
diagram is expected to look qualitatively as sketched in Fig.8b [\mikeref].
Note that contrary to the single flavour case, the prediction of this phase
structure in the $M-M_5$ plane is quite nontrivial.

Guided by this simple picture at weak coupling, I set out to look
for these phase transitions in the Hartree-Fock approximation. For numerical
work, I again fixed the values of the hopping parameter $(K=1)$ and
the Wilson parameter $(r=0.5)$,
so that the critical value of $M$ at zero coupling is
$M_{c}(0)=1$. Also, I have always included a very small explicit flavour
breaking in both $M$ and $M_5$. In particular, the masses of flavours
were of the form $M\pm 0.001$ and $M_5\pm 0.001$. The self-consistent
solutions were again obtained by the modified iteration procedure as
described in Appendix.
For all results presented here the accuracy of the solution,
given by Eq.~(\error), was better than $10^{-3}$ and typically about
$10^{-6}$. To achieve this accuracy in the immediate vicinity of the phase
transition on a lattice of $40$ sites took less then $10^4$ iterations
at strongest coupling studied here $(g^2=6)$.

The graphs in Figs.9a,b illustrate how the above qualitative ideas are
reflected in the Hartree-Fock approximation. In Fig.9a I plot the vacuum
expectation value of the electric field along the $M_5$-axis ($M=0$)
at $g^2=0.1$ on a lattice with $40$ sites. Note that the H-F vacuum nicely
exhibits the expected abrupt change in the electric field and the reversal
of its sign. Probing the field along the $M$-axis (with small $M_5$ present)
gives the dependence ploted in Fig.9b. While behaving in qualitatively
the same way, the magnitude of the field is becoming small as one approaches
the $M-g^2$ plane. This is what one would expect if these $\theta =\pi$
transitions, which are naturally first order, end in a second order
chiral endpoint in this plane. The anlysis of the electric field
correlators suggests however, that in the H-F approximation, these
transitions at finite coupling are first order even close to the
$M-g^2$ plane. It should be stressed again however, that this might well
be an artifact of the approximation.

Defining the transition point as the position in the $M-M_5$ plane where the
field switches its sign, I plot the phase diagrams for $g^2=0.1$ and
$g^2=1.2$ on a lattice with $32$ sites in Fig.10. In fact, only the points
in the upper right quadrant were really calculated. The rest of them were
obtained using symmetry with respect to the mass reflections. Note that while
not exactly of circular shape, the transition lines reflect the expected
qualitative features deduced from the surface mode picture. Also,
at $g^2=0.1$, non-negligible finite-size effects are present here.
While the transition point along the $M$-axis is essentially
stable against the increase of the lattice size, the transitions along the
$M_5$-axis occur at $M_5=0.36,0.42,0.48$ on the lattices with $32,40$ and
$48$ sites respectively. Thus it is quite possible that the ellipse-like
shape of the transition line will become more circle-like
in the infinite volume limit.

Let me now turn to the question of Aoki's phase in the $M-g^2$ plane. First
note that the surface mode picture doesn't suggest that parity-flavour broken
phase should exist here at weak coupling. Indeed, as one turns off the $M_5$
and moves along the $M$-axis, the angle $\theta$ defined
by (\thedef) is always zero. Both flavours will
generate their surface mode as $M$ is lowered below $M_{c}$, thus
changing $\theta$ by $2\pi$ and physically changing nothing.
A good way to picture this is by looking at Fig.8b: However small the flavour
breaking is, if one moves close enough to the $M$-axis, it is always
possible to pass below the chiral point.

Discussion of these issues in two dimensions might appear a little academic
since because of the Mermin-Wagner theorem [\refname\mewaref], one would not
expect flavour to be spontaneously broken here at {\it any} coupling.
Nevertheless, if the surface mode picture is of relevance to QCD in four
dimensions, then this qualitative prediction wouldn't change.
Moreover, I consider the following to be good reasons
to investigate these issues in the context of the Schwinger model itself:
1) The lattice model, defined by (\twoflham2), is nonlocal, and as such
does not exactly satisfy the usual assumptions of the theorem.
2) Parity can still be broken. 3) Flavour could be erroneously broken within
the H-F approximation. Then, although not useful as an information about
the lattice Schwinger model, it can serve as a toy picture of what might be
happening in QCD where there is quite convincing evidence that parity-flavour
is broken at strong enough coupling.

With that in mind, I have calculated the expectation values of
$\psibar\gamma_5\tau_3\psi$ and $\psibar\gamma_5 1\psi$ in the
H-F vacua. Here $\tau_3$ is the third Pauli matrix and $1$ a unit
matrix in flavour space. Note that the form of flavour breaking
in $M_5$ used here chooses the $\tau_3$-direction if flavour is broken.
Note also that if $M_5=0$ and $\langle\psibar\gamma_5\tau_3\psi\rangle\ne 0$,
it is both parity and flavour that are spontaneously broken. On the other
hand, if this expectation is nonzero at nonzero $M_5$, the parity is broken
explicitly while the flavour spontaneously. Furthermore, if at $M_5=0$
we had $\langle\psibar\gamma_5 1\psi\rangle\ne 0$ and
$\langle\psibar\gamma_5\tau_3\psi\rangle =0$, it would indicate that
only parity has been spontaneously broken.

The results of the H-F analysis in $M-g^2$ plane (with $M_5=10^{-2}$)
on the lattices with up to $40$ sites are as follows. I have found no
evidence of $\psibar\gamma_5 1\psi$ acquiring an expectation value
in the region of couplings $g^2\le 6$. Consequently, there is no
indication of parity being broken alone. However, there are regions
where $\langle\psibar\gamma_5 \tau_3\psi\rangle$ is nonzero
in the Hartree-Fock approximation on the finite lattice.
For example, in Fig.11a I plot this expectation as a function
of fermion mass at $g^2=1.2$. The broken region appears as a narrow peak
adjacent from the left to the ``$\theta=\pi$''
\footnote*{Note that I loosely refer to a transition, where electric field
switches the sign as the ``$\theta=\pi$'' transition even at strong coupling.
This should not be taken too literally neither here, nor in what follows.}
transition point on this lattice of
$40$ sites. I have observed similar peaks at $g^2=0.1$ and $g^2=4.0$ with
heights roughly $0.08$ and $0.70$ respectively.
For the two weaker of the above
couplings, I have also performed a finite-size analysis of the width
of this broken region. This width decreases linearly with $1/L$, exhibiting
a small negative intercept in both cases. On the other hand, the heights
of the peaks stay constant as the lattice size increases. I therefore
conclude that these narrow regions will not survive in the H-F phase
diagram in the infinite volume limit. The only remnant of them will
probably be the singular behaviour of the parity-flavour order parameter
at the ``$\theta=\pi$'' transition point.

The situation qualitatively changes at even stronger couplings.
In particular, the parity-flavour broken phase indeed opens up at the
subcritical fermion masses. This is demonstrated in Fig.11b, where
I show the behaviour of the order parameter at $g^2=6.0$ on a lattice
with $32$ sites. There has been a negligible change here as the lattice
size increased to $L=40$. I therefore expect the finite-size effects to be
small. It is also worth mentioning that
$\langle\psibar\gamma_5\tau_3\psi\rangle\ne 0$ in the whole
inside region of the ``$\theta=\pi$'' line at this
strong coupling and not only in the $M-g^2$ plane.
This is in contrast to the case of narrow broken regions at
weaker couplings which can only be observed close to $M-g^2$ plane
on a finite lattice.

This concludes the review of the most important aspects of the numerical
information obtained in this study. In the last section, I will turn
to generalizations and speculations.

\bigskip
\bigskip
{\bf VII. Summary, Generalizations and Speculations}
\bigskip
\bigskip

The Schwinger model on a lattice with Wilson fermions has been studied
in the Hartree-Fock approximation. The main focus
was given to the global structure of phase diagrams with one and
two degenerate flavours of fermions. In future communication, I plan
to report on the study of the continuum limit in this framework.
The surface mode picture [\mikeref,\ourref] served as a reliable guide
in these investigations at weak coupling. The nonperturbative nature
of H-F approximation however, allows to study the model at
intermediate and strong couplings as well.

For the case of a single flavour, I plot in Fig.12 the qualitative behaviour
of the concluded full phase diagram of the model in the Hartree-Fock
approximation. There is a planar region, embedded in the $M-g^2$ plane,
where parity is spontaneously broken. From the point of view of the surface
mode picture, it can be understood as the surface of ``$\theta=\pi$''
transitions. Entering the region from the $M$-direction is accompanied
by the appearance of the surface charges and the background electric
field (``$\theta=0 \rightarrow \theta=\pm\pi$'').
Crossing the region in the $M_5$-direction corresponds to reversing the sign
of the electric field (``$\theta=\pi \leftrightarrow \theta=-\pi$'').
In accordance with Aoki's scenario, $\psibar\gamma_5\psi$ acquires an
expectation value in the broken region. Taking into account the
infinite-coupling result of [\galaref], I expect the
parity-violating phase to extend all the way to $g^2\rightarrow\infty$.

While the ``$\theta=\pi \leftrightarrow \theta=-\pi$'' transitions
are naturally first order, there is a strong evidence that the
parity-violating phase transitions in $M-g^2$ plane
are also first order in the Hartree-Fock approximation at nonzero $g$.
This seems quite unnatural since one would have the first order ends
at the boundaries of the parity-violating region. On the other hand,
if one thinks conventionally about taking the chiral continuum limit,
the first order transition looks quite appropriate. There are two parts
to the conventional wisdom about taking this limit, which is usually thought
about in analogy to QCD in four dimensions. First, since the gauge
coupling constant is dimensionful (inverse length) in two dimensions,
it is assumed that the continuum limit can only be taken
at vanishing $g$ (the dimensionless lattice coupling). Second, the chiral
limit is assumed to be taken by following the line of phase
transitions $M_{c}(g^2)$ towards $g=0$. The immediate consequence of these
assumptions is that $g(a)$ vanishes at $a=0$ and is an increasing
function in the vicinity of this point. Here $a$ is the lattice spacing.
Denoting by $\Delta(a)$ the dimensionless mass gap along the line
of phase transitions, the physical mass of the lightest particle
in the theory is given by $\Delta(a)/a$.
Since chiral symmetry is broken in the continuum and the lowest mass
is nonzero $(e/\sqrt\pi)$, the above ratio should approach the constant
positive value as $a\rightarrow 0$. Consequently,
similarly to $g(a)$, $\Delta(a)$ should also vanish at $a=0$ and increase
in the vicinity of this point. Puting the above two conclusions
together, $\Delta(g)$ must have this local property as well.
In particular, it is zero at $g=0$, but increases as $g$ becomes finite.
Hence, if the conventional picture about chiral continuum limit is correct,
the transitions should become first order as the gauge coupling is turned on.

In the light of the above considerations, it is not entirely obvious
that the H-F approximation is giving an incorrect answer
here (which it of course well can). One possible solution is that
what is depicted in Fig.12 is not all that happens in the model.
In particular, there could be another sheet of first order phase
transitions going off the $M-g^2$ plane and crossing this plane at
$M_{c}(g^2)$. If that was the case, then the second order ends at
$M_{c}(g^2)$ would not appear to be necessary. However, I have
not found the evidence that would support this scenario in the H-F
approximation. In summary, the order of parity violating phase
transition is a very interesting issue by itself. However, it can
only be satisfactorily settled by accurate calculation
beyond the H-F approximation.

Similarly to the one-flavour case, there is a surface of ``$\theta=\pi$''
transitions also in the model with two flavours. It takes a more complicated
shape however and its qualitative behaviour in the H-F approximation
is depicted in Fig.13a. The ``tube'' of phase transitions encloses
the $g^2$-axis as the explicit flavour breaking is taken to zero and touches
the $M-g^2$ plane at $M_{c}(g^2)$. Since the electric field generated
by the surface charges switches the sign, the transitions across the surface
of the ``tube'' are naturally first order. In the H-F approximation, this is
so even when approaching $M_{c}(g^2)$ at finite coupling. The transition
close to $M_c(g^2)$ becomes second order in H-F approximation
only as $g\rightarrow 0$. Note also that contrary to the
single-flavour case, the possible continuous nature of the phase
transitions along $M_{c}(g^2)$ would not be in conflict with the above
argument concerning the chiral continuum limit. This is because in the
 multi-flavour case, the nonsinglet part of the flavoured chiral symmetry
is not anomalous and there is a massless particle in the continuum theory.

The crucial difference between Fig.12 and Fig.13a is that with two flavours,
it is only $M_{c}(g^2)$ that is shared by the $M-g^2$ plane and the surface
of ``$\theta=\pi$'' transitions. If the identification of the
``$M_5$''-physics on the lattice and the ``$\theta$''-physics in the
continuum is correct at weak coupling, nothing special should happen upon
crossing $M_{c}(g^2)$ with respect to parity, and it indeed doesn't.
At strong coupling however, the above scenario might well break.
This is nicely observed in the H-F approximation.
In particular, the numerical evidence suggests the existence of rather
strong coupling $g_s$ ($4<g_s^2<6$), so that for $g>g_s$, the expectation
value of $\psibar\gamma_5\tau_3\psi$ is nonzero inside the ``tube'' of
Fig.13a. Consequently, parity-flavour is broken in $M-g^2$ plane at
subcritical masses and strong couplings.
I conclude the qualitative H-F phase diagram in this
plane as depicted in Fig.13b. The full lines in this phase diagram represent
$M_{c}(g^2)$ and they are also characterized by the fact
that $\langle\psibar\psi\rangle$ exhibits a jump as they are crossed.
Parity-flavour however, is only broken in the ``BP''-region,
bounded from bellow by the dashed line.

The above results indicate that Aoki's scenario is not realized in the
two-flavour Schwinger model at the H-F level. It is quite feasible,
that this is the case for QCD in four dimensions
as well [\refname\mikenewref,\refname\bitarref].
Although the analogy between $QED_2$ and $QCD_4$ should certainly
not be taken too seriously (especially in case of an approximation),
I believe that the phase diagram of Fig.13b indeed represents
a possible toy picture of what might be happening in the latter case.
In particular, that the parity-flavour broken phase shrinks
to zero width before entering the vicinity of the continuum limit.
There would still be a line $\kappa_c(g^2)$, running up from
the QCD fixed point, on which a transition
in $\langle\psibar\psi\rangle$ could be observed. However,
$\langle\psibar\gamma_5\tau_3\psi\rangle$ would remain zero.

\bigskip
\bigskip
\noindent {\bf Appendix}
\bigskip
\bigskip

In this appendix I will briefly describe a technical detail on
the numerical procedure used to solve the H-F equations (\hafoeq).
The standard way to proceed is to
iteratively generate the sequence of sets of one particle states
$\{S^{(0)}, S^{(1)},\ldots S^{(k)}, \ldots\}$, so that $S^{(k+1)}$ is the
eigenset of $\hhf^{(k)}$. Here $\hhf^{(k)}$ is the H-F Hamiltonian
with direct and exchange potentials determined from wavefunctions
of $S^{(k)}$. With a reasonable choice of the initial set the
sequence frequently converges well to the self-consistent set
$S_{HF}$.

Note however, that there is certain ambiguity in the procedure
that might cause a problem. It arises because of the fact that
the energy of the H-F vacuum is not just a sum of the one particle
energies of the filled levels. Indeed, at nonzero coupling
there is an interaction part contributing to the total energy and
it may well be that the filled levels are not those from $S^{HF}$
with lowest one-particle energies. If that is the case and the
iteration proceeds by filling the lowest levels at each step,
the procedure can never converge to a self-consistent set.

This bad looking flaw can however be quite easily rectified
[\negeleref]. Instead of the original H-F problem (\energy,\hafoeq),
consider the one with the two body potential shifted
by a constant, i.e.
$$
\mid n-m\mid\quad \longrightarrow\quad \mid n-m\mid +\, C \,.
\neweq
$$
One naturally expects that a resulting H-F vacuum will not be
physically different from that of the original problem.
Indeed, it can be easily checked that both problems share
their solutions. However the vacuum energy and also the one
particle energies will change. In particular,
$$
\eqalign{E \quad&\longrightarrow\quad E \,+\, CL{g^2\over 4}\cr
         \epsilon^\alpha \quad&\longrightarrow\quad \cases{
         \epsilon^\alpha + C{g^2\over 2}, &$\alpha$ filled;\cr
         \epsilon^\alpha, &$\alpha$ empty.}\cr}
\eqname{\shifts}
$$

Note that it is only the filled levels that get shifted in energy,
not the empty ones.
Therefore by choosing $C$ to be negative and sufficiently large one
can allways make the filled levels to be those with lowest one-particle
energies. The H-F problem with such $C$ can then in principle be
solved by standard iteration as described above. If the solution
is found, it is also the H-F vacuum of the original problem.

In an actual computation, the constant $C$ was chosen by trial and error.
If the iteration failed for a given $C$, a larger value has
been set. In general, larger values were needed for larger values
of $g$, as one would expect.

\vfill\eject

\noindent {\bf Acknowledgements}
\bigskip
\bigskip
I am indebted to Mike Creutz for many valuable discussions
and for helping me to rectify some of the lingual shortcomings
in the preliminary version of this paper.
I have benefited from discussions with Tony Kennedy, Urs Heller,
Khalil Bitar, Claudio Rebbi and Jorge Piekarewicz. Thanks also
to John Negele for bringing the Ref.~[\negeleref] into my attention.
The financial support from DOE under Grant Nos. DE-FG05-85ER250000 and
DEFG05-92ER40742 is also acknowledged.

\bigskip
\bigskip
\noindent{\bf References}
\bigskip
\bigskip

\item{\nogoref.} H.~Nielsen, M.~Ninomiya, Nucl.~Phys.~B185 (1981) 20;
                 Nucl.~Phys.~B193 (1981) 173.

\item{\mikeref.} M.~Creutz, Nucl.~Phys.~B (Proc.~Supl.) 42, 56 (1995).

\item{\shamref.} Y.~Shamir, TUAP-2287-95, hep-lat/9509023.

\item{\currentalgebraref.} S. Treiman, R. Jackiw, B. Zumino, and E. Witten,
                           {\sl Current algebra and anomalies,}
                           (World Scientific, 1985, QC793.3.A4C87).

\item{\anomref.} L.~Karsten, J.~Smit, Nucl.~Phys.~B183 (1981) 103;
                 A.~Coste, C.~Korthals-Althes, O.~Napoly,
                 Phys.~Lett.~B179 (1986) 125.

\item{\aokiref.}  S.~Aoki, Phys.~Rev.~Lett. 57 (1986) 3136;
                            Nucl.~Phys.~B314 (1989) 79.

\item {\aogoref.}  S.~Aoki, A.~Gocksch, Phys.~Rev.~D45 (1992) 3845.

\item {\aoukumref.}  S.~Aoki, A.~Ukawa and T.~Umemura,
                     UTHEP-313, Aug 1995; hep-lat/9508008.

\item{\abgref.} S.~Aoki, S.~Boettcher, A.~Gocksch,
                Phys.~Lett.~B331 (1994) 157.

\item{\bivrref.} K.~Bitar, P.~Vranas Phys.~Rev.~D 50 (1994) 3406.

\item{\kaplanref.}  D.~Kaplan, Phys.~Lett.~B288 (1992) 342.

\item{\jansenref.} K.~Jansen, DESY-94-188, hep-lat/9410018.

\item{\ourref.} M.~Creutz, I.~Horv\'ath Phys.~Rev.~D 50 (1994) 2297.

\item {\colemanref.} S.~Coleman, Ann.~Phys. 101 (1976) 239.

\item {\schconref.} C.~R.~Gattringer, E.~Seiler, Ann.~Phys. 233 (1994) 97;
            J.~E.~Hetrick, Y.~Hosotani, S.~Iso, Phys.~Lett.~B350 (1995) 92.

\item {\ksschw.} E.~Marinari, G.~Parisi, C.~Rebbi,
                 Nucl.~Phys.~B190 [FS3] (1981) 734;
                 S.~R.~Carson, R.~D.~Kenway, Ann.~Phys. 166 (1986) 364.

\item {\galaref.} H.~Gausterer, C.~H.~Lang, Phys.~Lett.~B341 (1994) 46.

\item {\azcoitiref.} V.~Azcoiti et al., INFN-LNF 94/009(P); hep-lat/9401032.

\item {\negeleref.} L.~L.~Salcedo, S.~Levitt, J.~W.~Negele,
                    Nucl.~Phys.~B361 (1991) 585.

\item{\mewaref.} N.~D.~Mermin, H.~Wagner Phys.~Rev.~Lett. 17 (1966) 1133.

\item{\mikenewref.} M.~Creutz, BNL-62123, May 1995; hep-th/9505112.

\item{\bitarref.} K.~Bitar, private communication.

\vfill\eject

\midinsert
\epsfxsize=1.0\hsize
\centerline{\epsffile {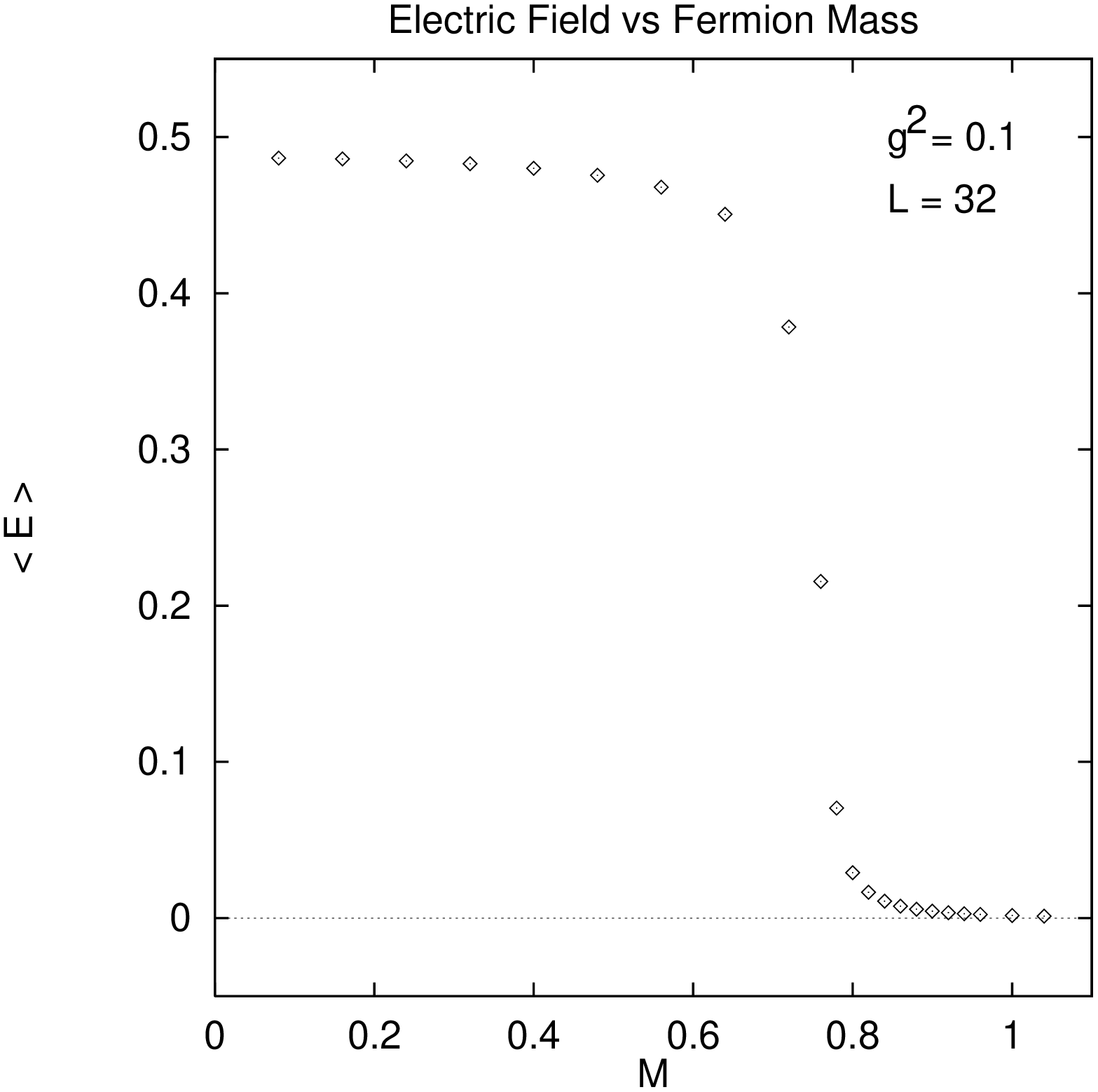}}
\vskip 1.05in
\noindent \narrower {{\bf Fig.1}
Vacuum expectation value of the electric field in the middle
of the system as a function of fermion mass.}
\endinsert

\vfill\eject

\midinsert
\epsfxsize=0.60\hsize
\epsfysize=0.40\vsize
\centerline{\epsffile {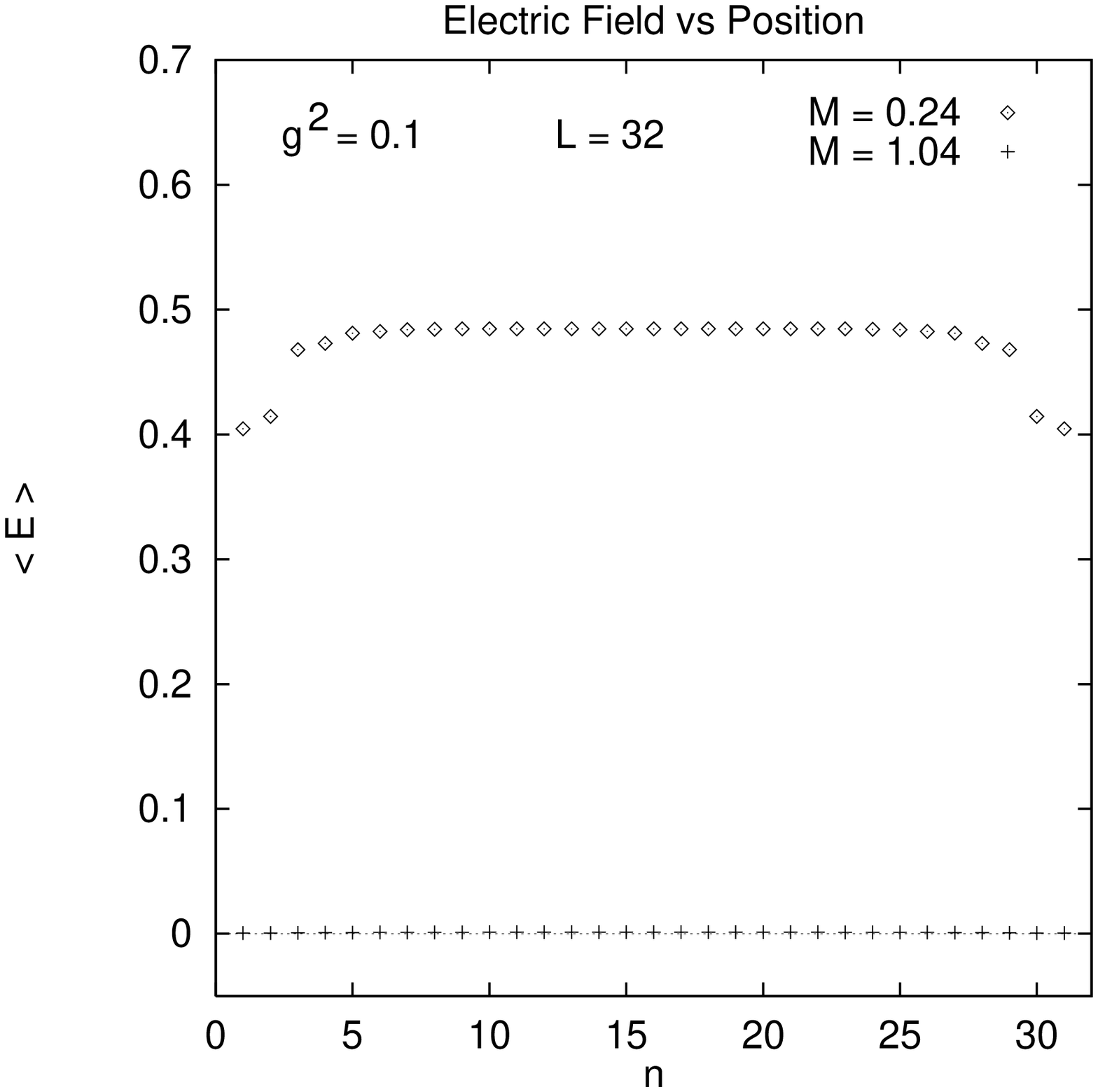}}
\endinsert

\vskip 0.08in
\midinsert
\epsfxsize=0.60\hsize
\epsfysize=0.40\vsize
\centerline{\epsffile {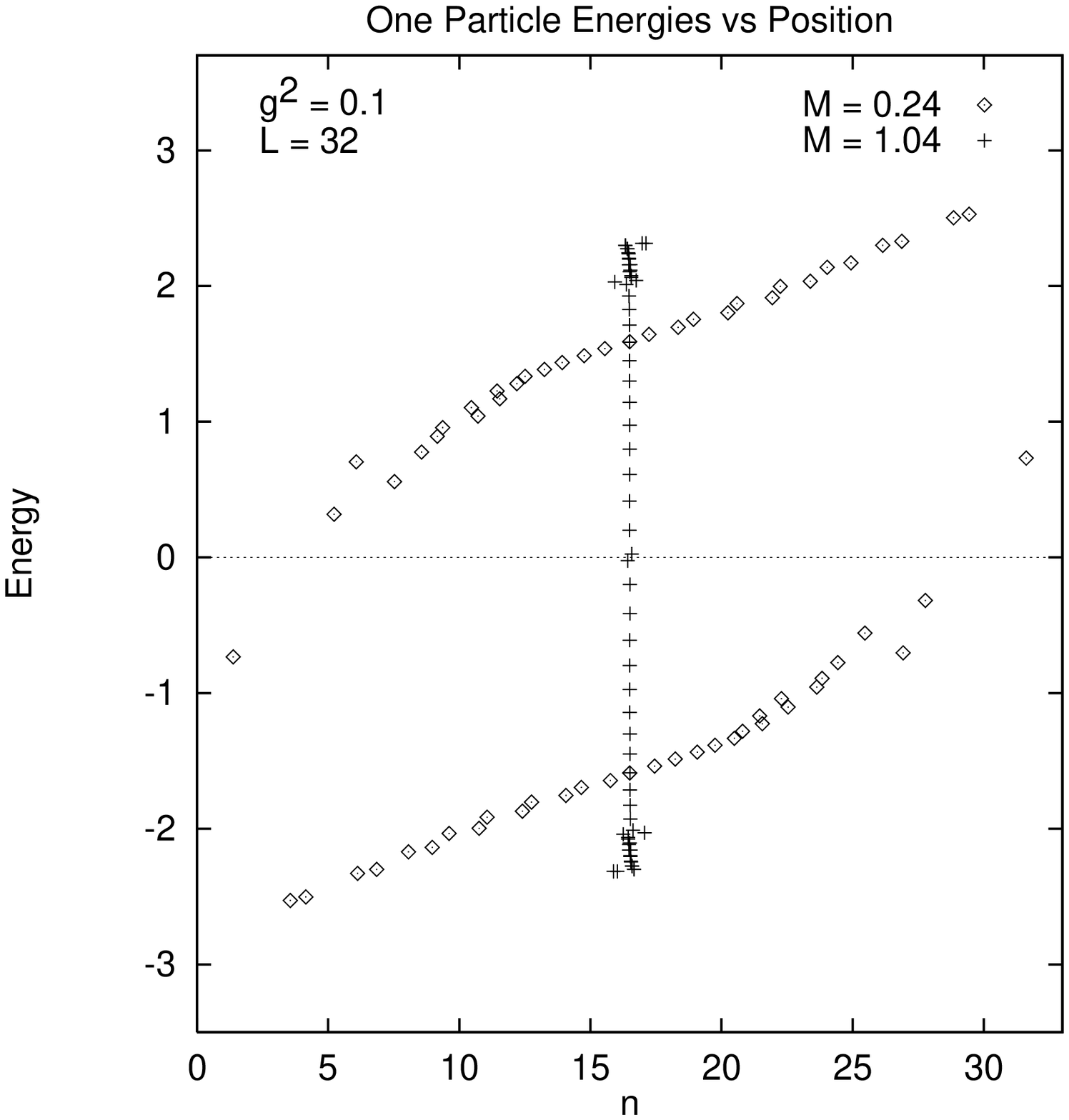}}
\vskip -0.20in
\noindent \narrower {{\bf Fig.2a} (Upper) Spatial dependence
of the electric field in symmetric $(M=1.04)$ and broken
$(M=0.24)$ phases. {\bf Fig.2b} (Lower) Spatial distribution
of the corresponding H-F levels.}
\endinsert

\vfill\eject

\midinsert
\epsfxsize=1.0\hsize
\centerline{\epsffile {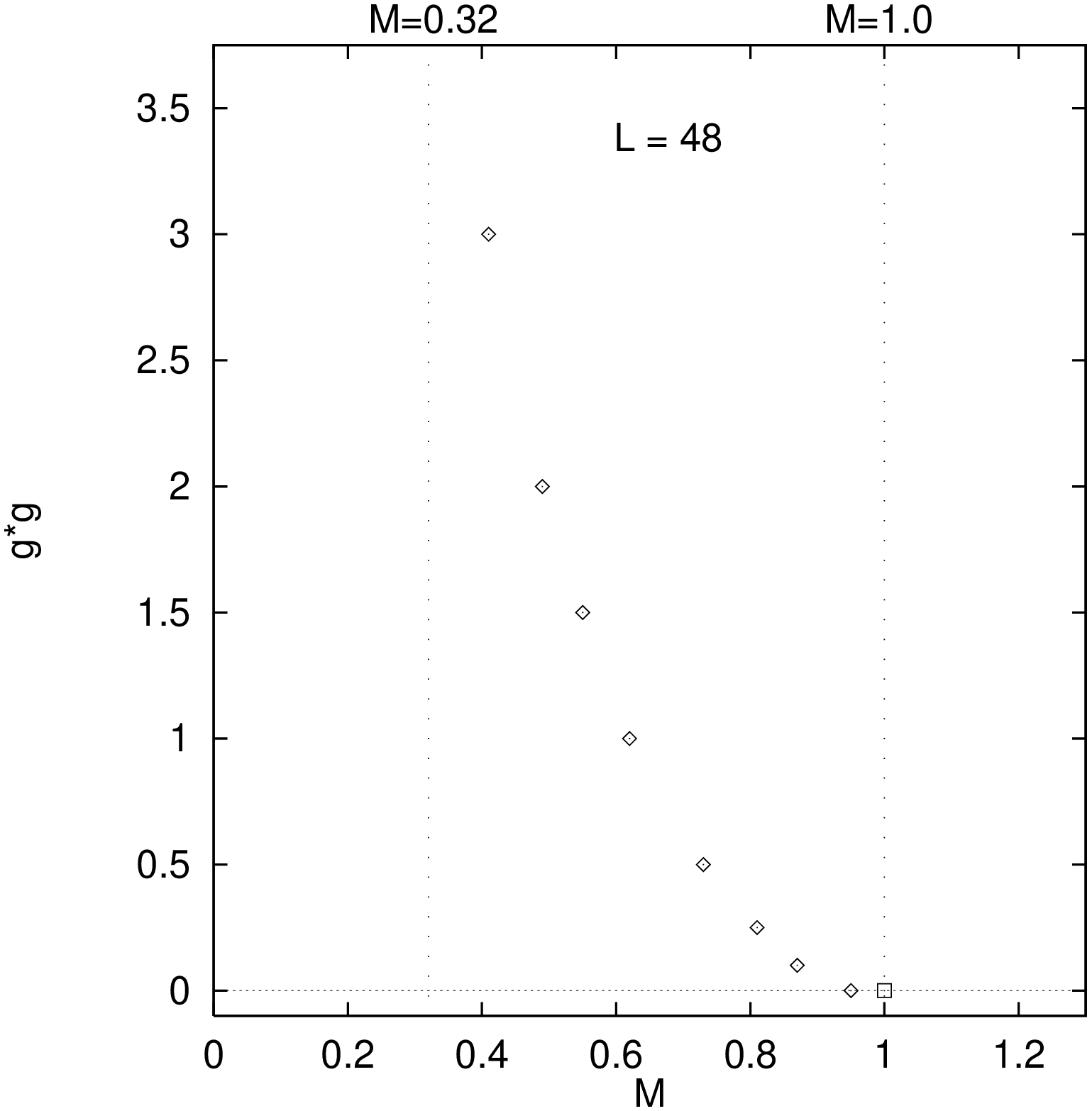}}
\vskip 1.0in
\noindent \narrower {{\bf Fig.3} Hartree-Fock phase diagram
in $M-g^2$ plane as seen on a finite lattice. Diamonds
sample the critical line $M_{c}(g^2)$, with parity broken
in the left region.
The square marks $M_{c}(0)$ at infinite volume limit.
The left vertical line represents $M_{c}(\infty)$ quoted
in [\galaref] for the model in standard Lagrangian formulation.}
\endinsert

\vfill\eject

\midinsert
\epsfxsize=0.60\hsize
\epsfysize=0.40\vsize
\centerline{\epsffile {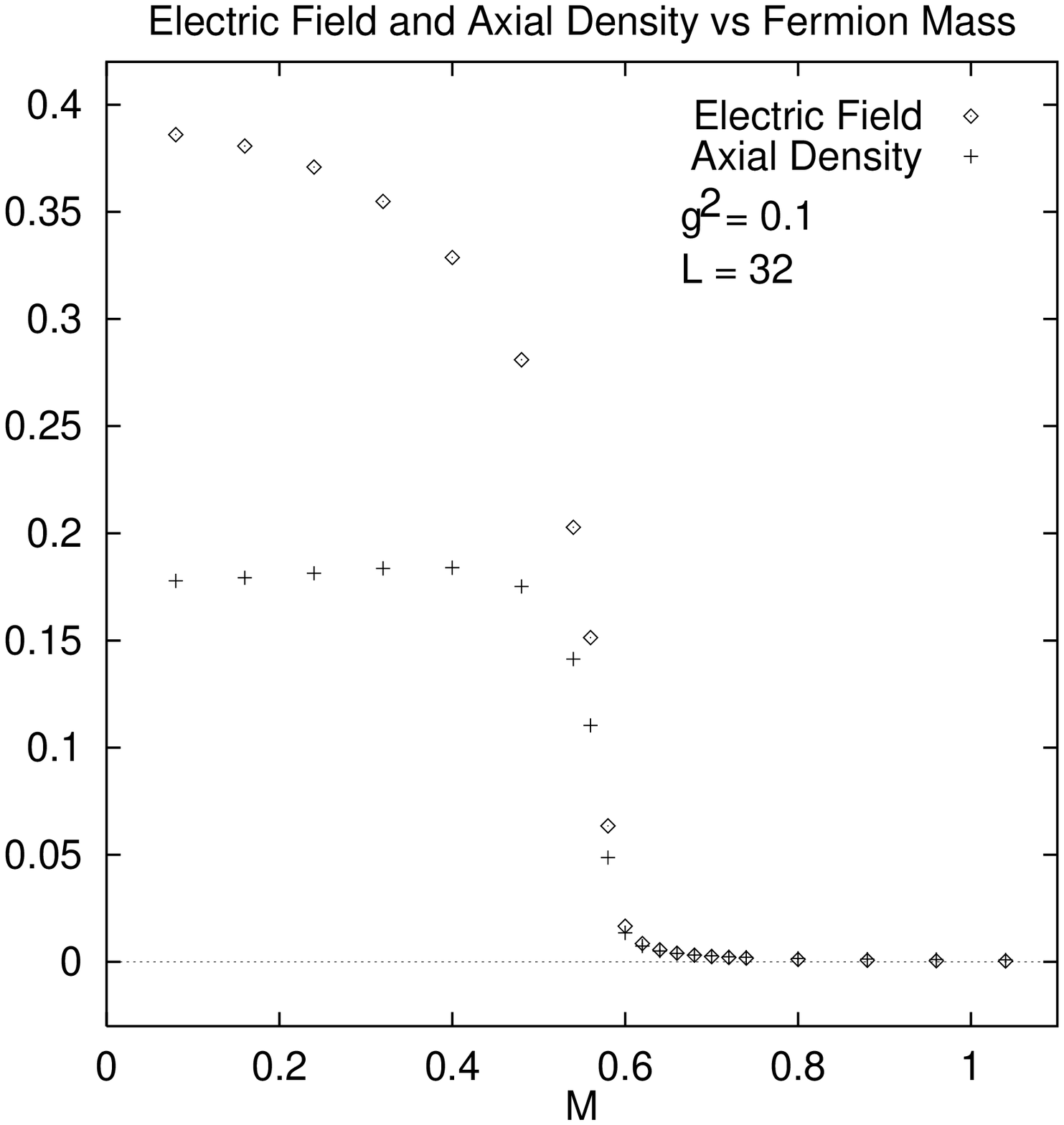}}
\endinsert

\vskip 0.08in
\midinsert
\epsfxsize=0.60\hsize
\epsfysize=0.40\vsize
\centerline{\epsffile {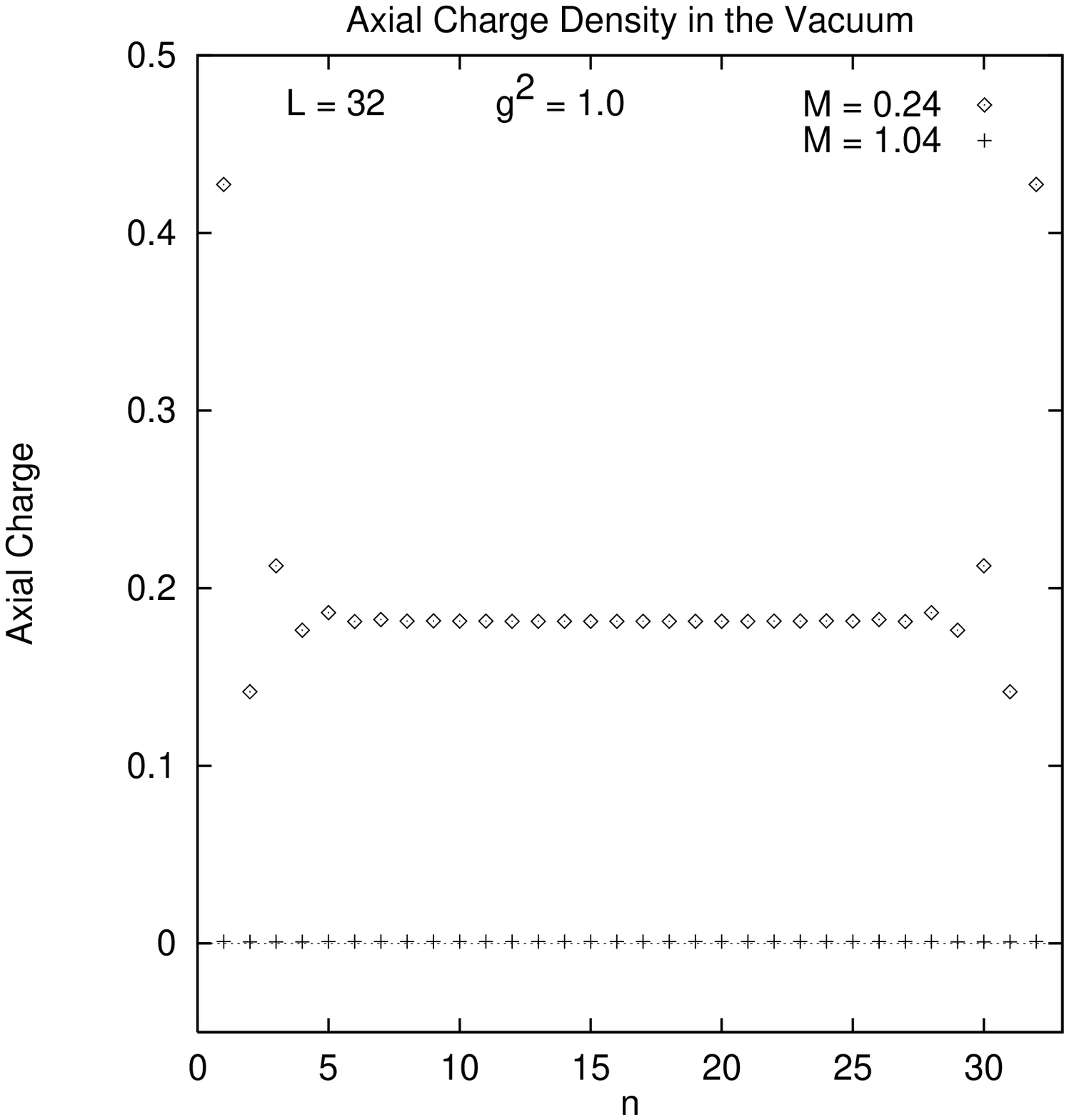}}
\vskip -0.20in
\noindent \narrower {{\bf Fig.4a} (Upper) Electric field and
axial charge density against the fermion mass. Transitions
seem to occur simultaneously.
{\bf Fig.4b} (Lower) Spatial dependence of axial density
in symmetric $(M=1.04)$ and broken $(M=0.24)$ phases.}
\endinsert

\vfill\eject

\midinsert
\epsfxsize=1.0\hsize
\centerline{\epsffile {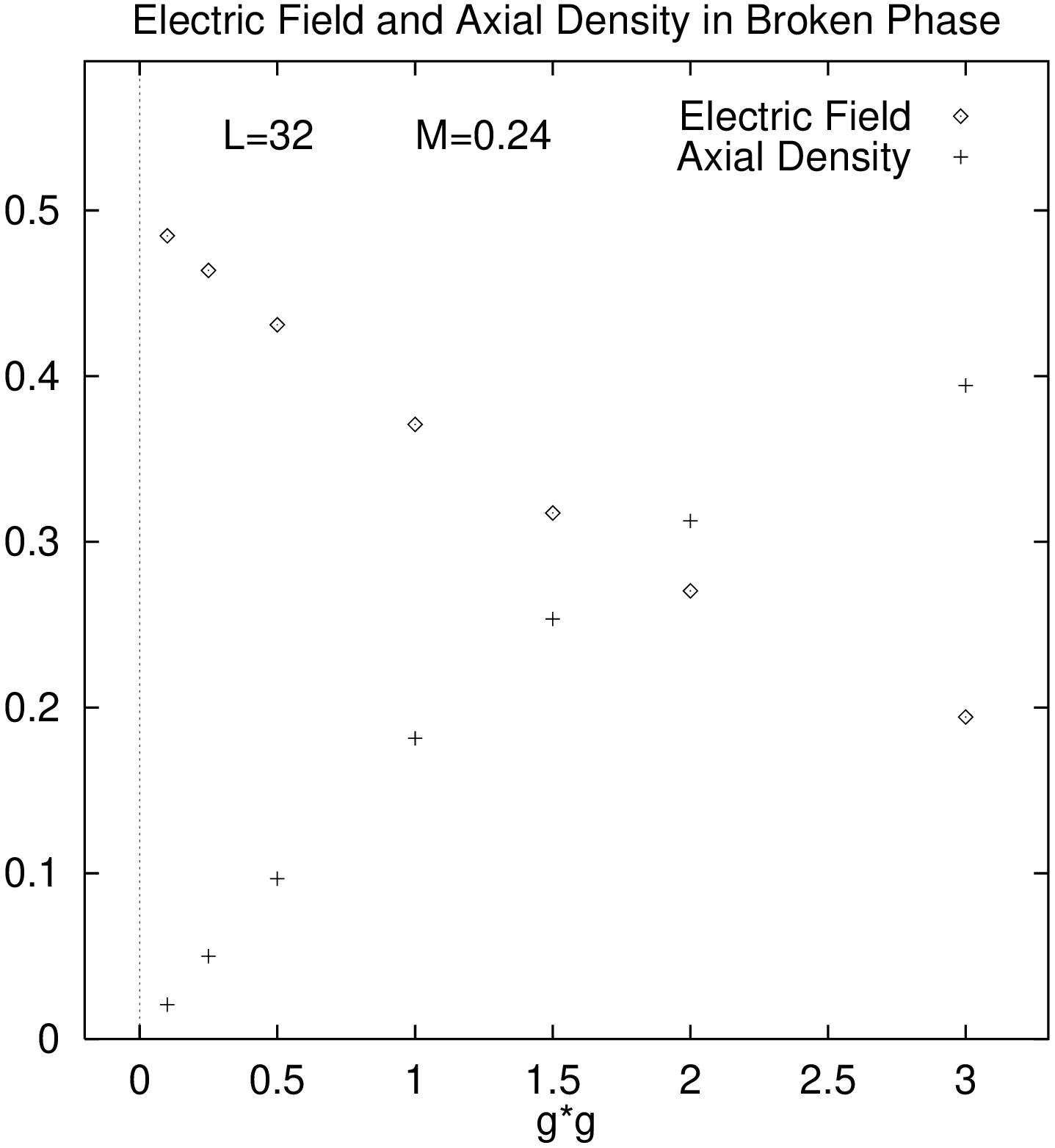}}
\vskip 1.05in
\noindent \narrower {{\bf Fig.5}
Expectation values of electric field and axial charge density
in broken phase. Relative size of the condensates varies with
gauge coupling.}
\endinsert

\vfill\eject

\midinsert
\epsfxsize=1.0\hsize
\centerline{\epsffile {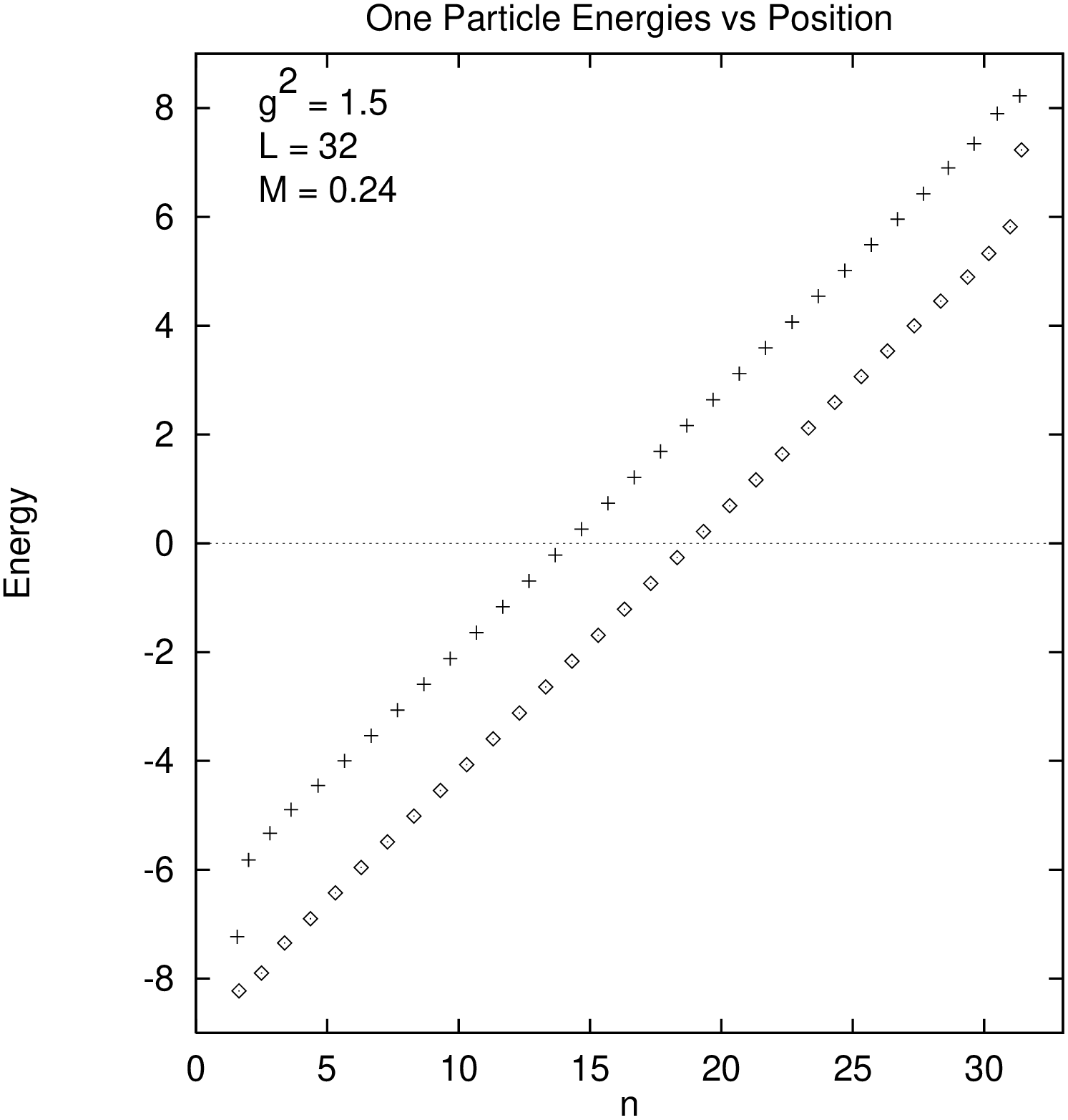}}
\vskip 1.05in
\noindent \narrower {{\bf Fig.6}
Spatial distribution of the H-F levels in broken phase at $g^2=1.5$.
Filled levels are marked by the diamonds and empty ones by the crosses.}
\endinsert

\vfill\eject

\midinsert
\epsfxsize=0.60\hsize
\epsfysize=0.40\vsize
\centerline{\epsffile {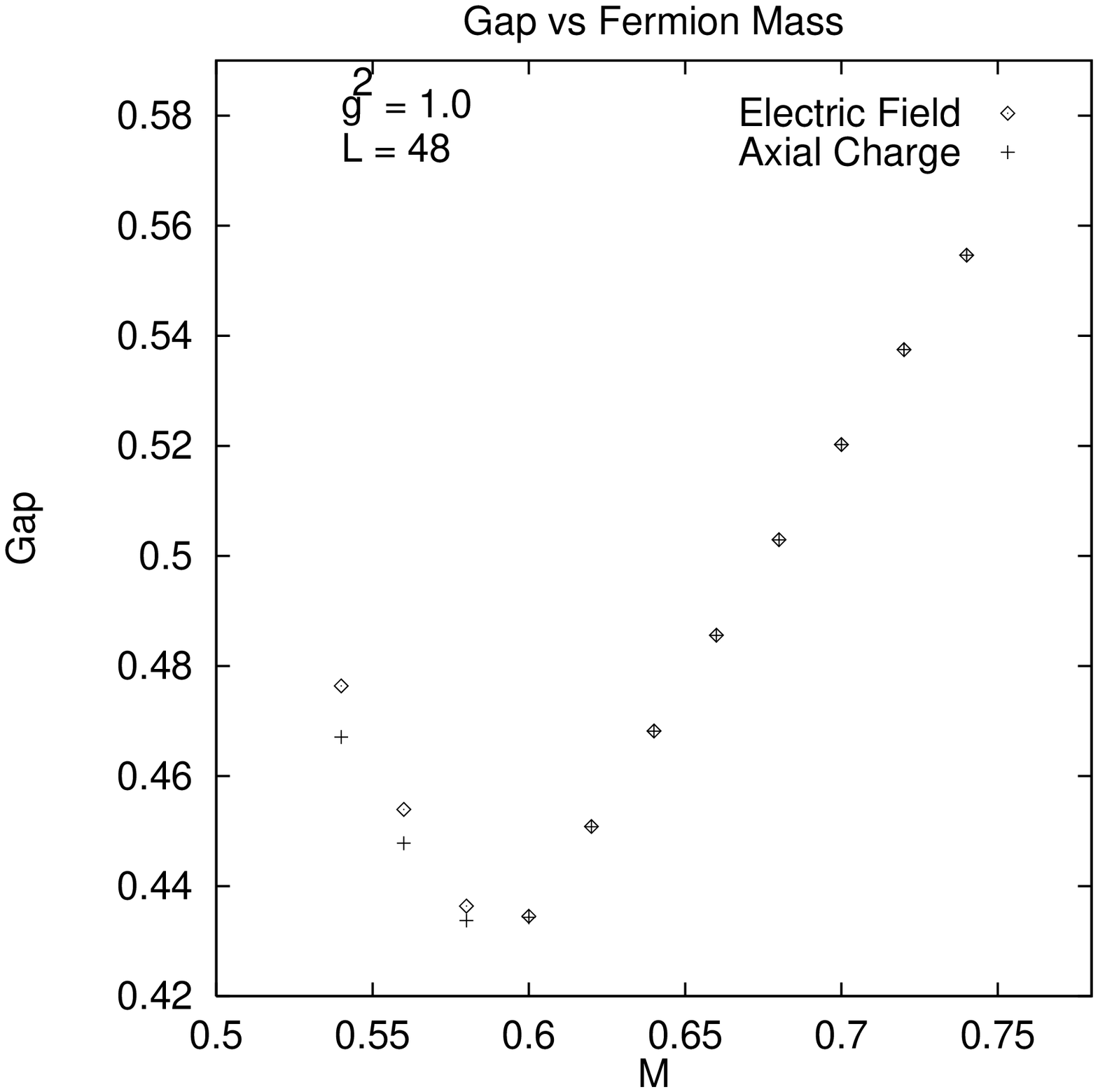}}
\endinsert

\vskip 0.08in
\midinsert
\epsfxsize=0.60\hsize
\epsfysize=0.40\vsize
\centerline{\epsffile {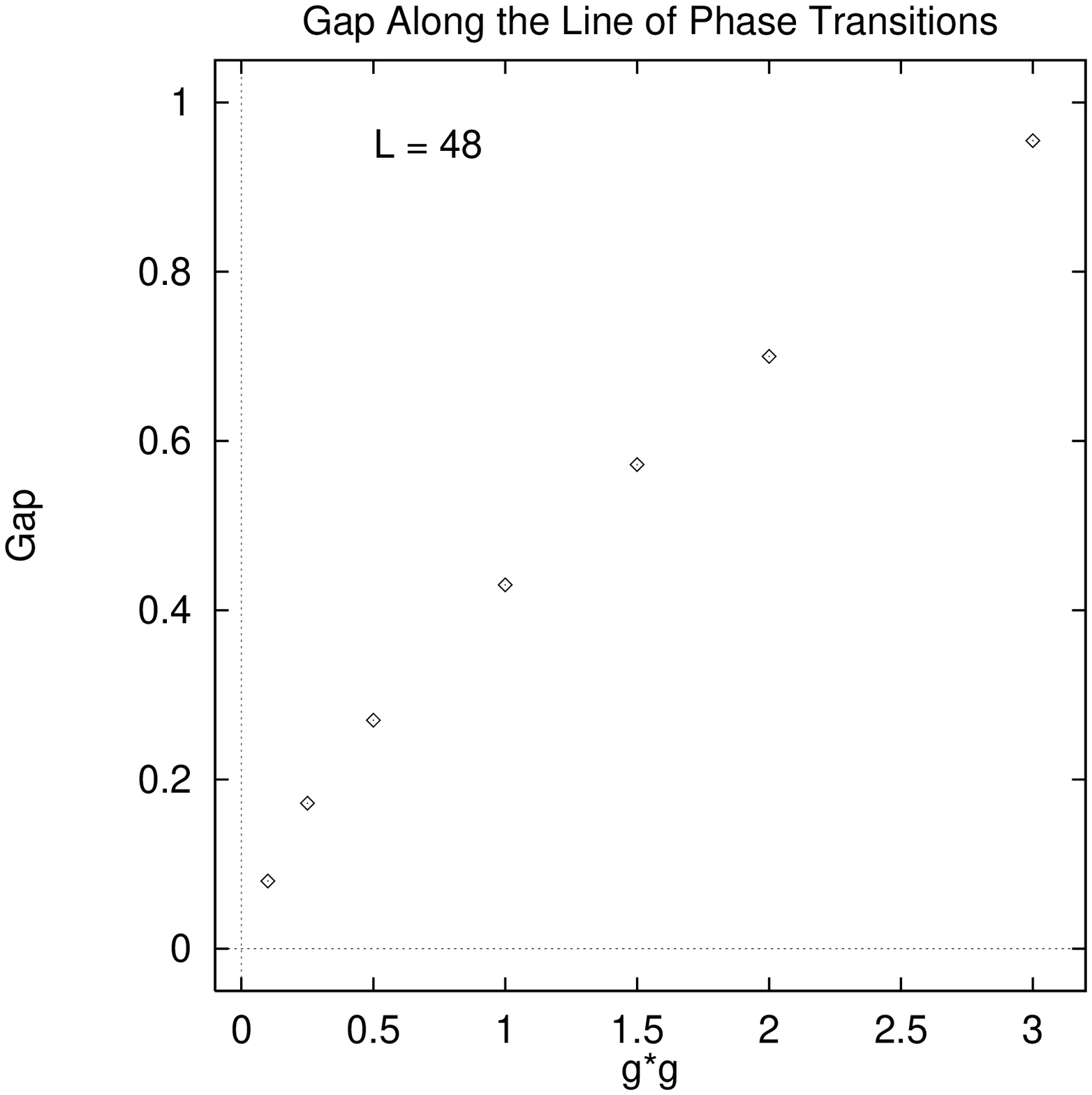}}
\vskip -0.20in
\noindent \narrower {{\bf Fig.7a} (Upper) Inverse correlation
length from electric field and axial charge correlation functions.
{\bf Fig.7b} (Lower) Inverse correlation length of the electric
field correlator along the line of phase transitions.}
\endinsert

\vfill\eject

\midinsert
\epsfxsize=0.60\hsize
\epsfysize=0.40\vsize
\centerline{\epsffile {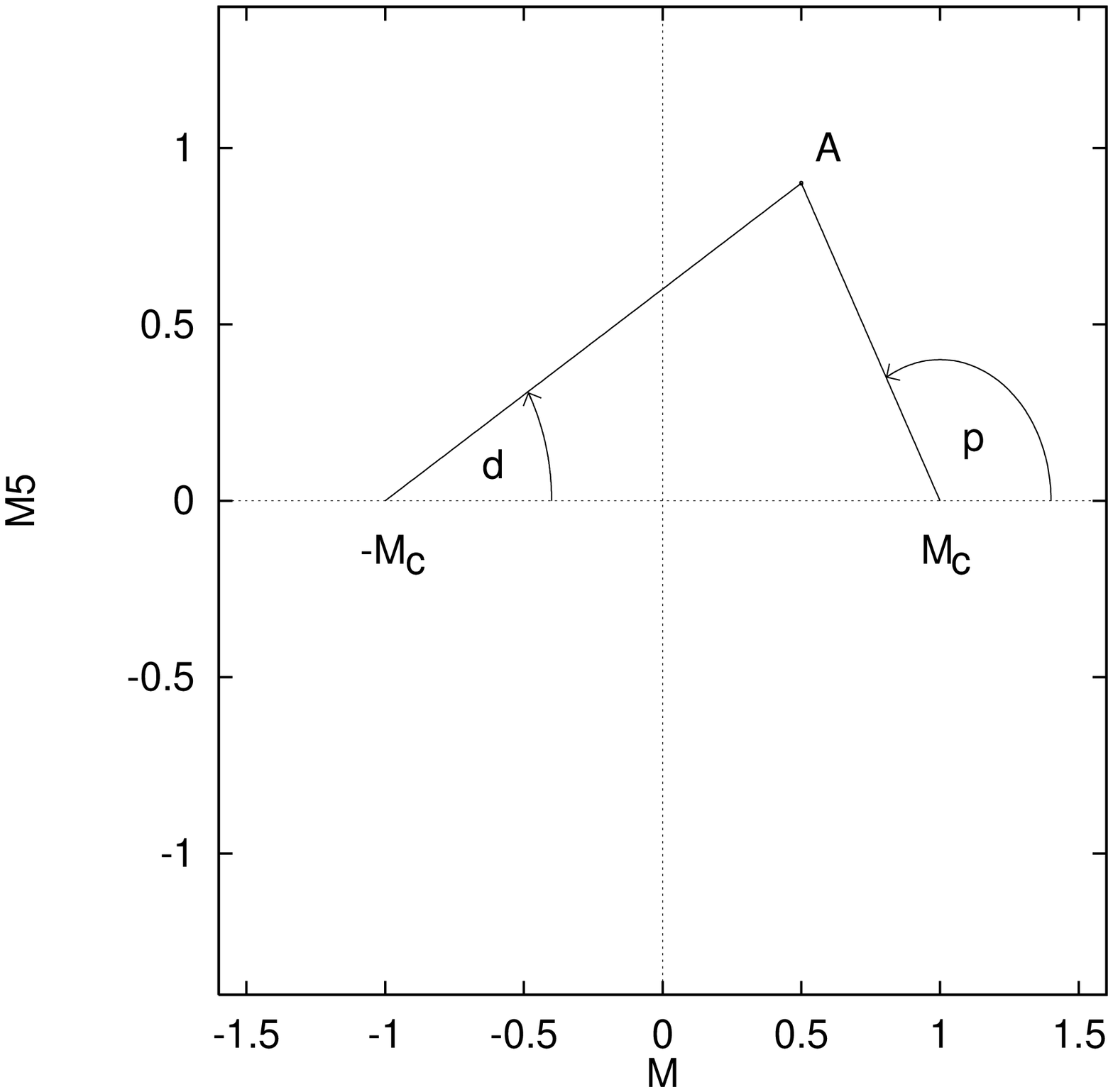}}
\endinsert

\vskip 0.08in
\midinsert
\epsfxsize=0.60\hsize
\epsfysize=0.40\vsize
\centerline{\epsffile {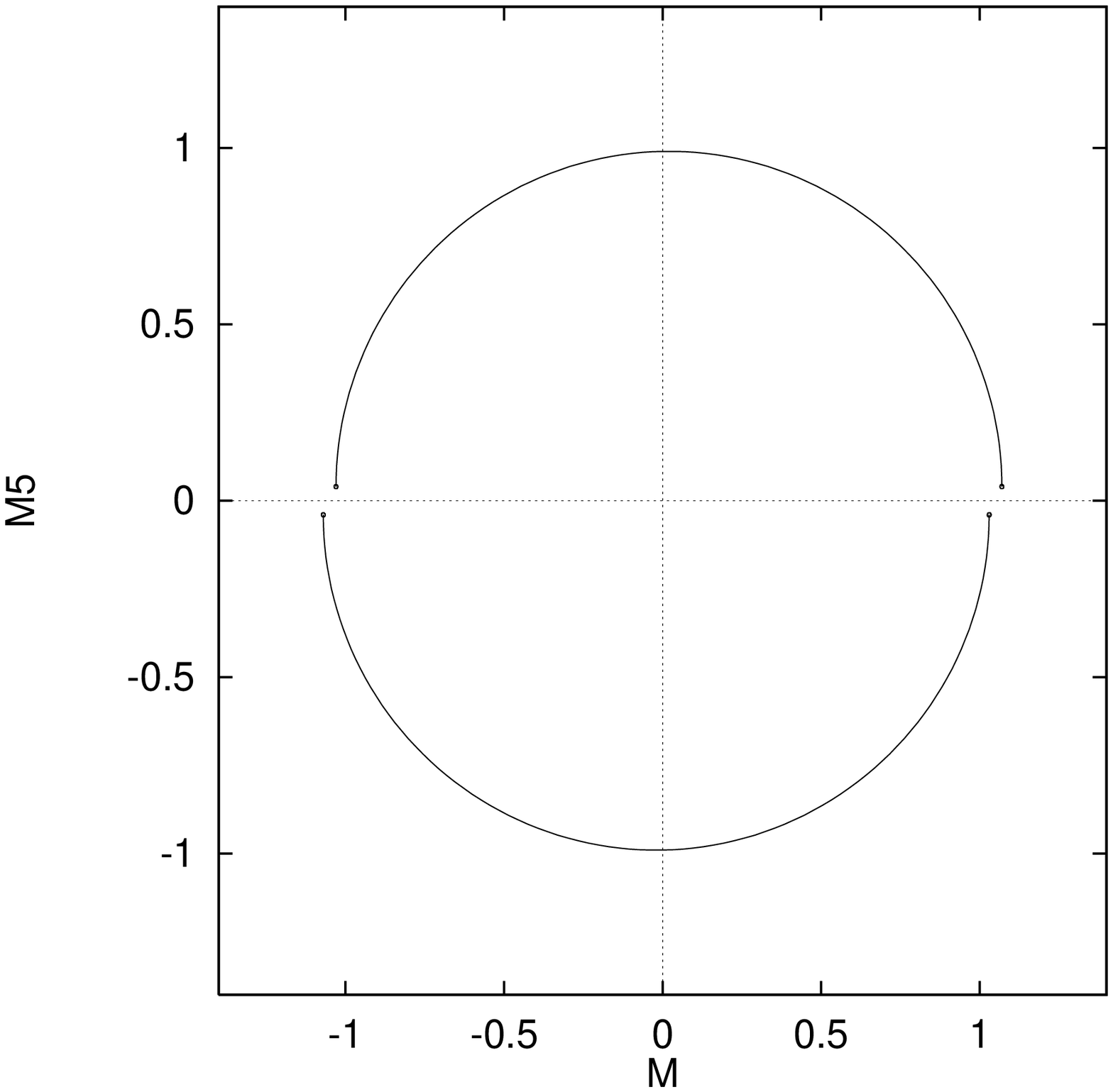}}
\vskip -0.20in
\noindent \narrower {{\bf Fig.8a} (Upper) Assignements of the
angles $\theta_p$ (p) and $\theta_d$ (d) of Eq.~(\thedef) to a general
point $A$ in the $M-M_5$ plane.
{\bf Fig.8b} (Lower) Expected phase diagram for two flavours at weak
coupling based on the surface mode picture of Ref.~[\mikeref].}
\endinsert

\vfill\eject

\midinsert
\epsfxsize=0.60\hsize
\epsfysize=0.40\vsize
\centerline{\epsffile {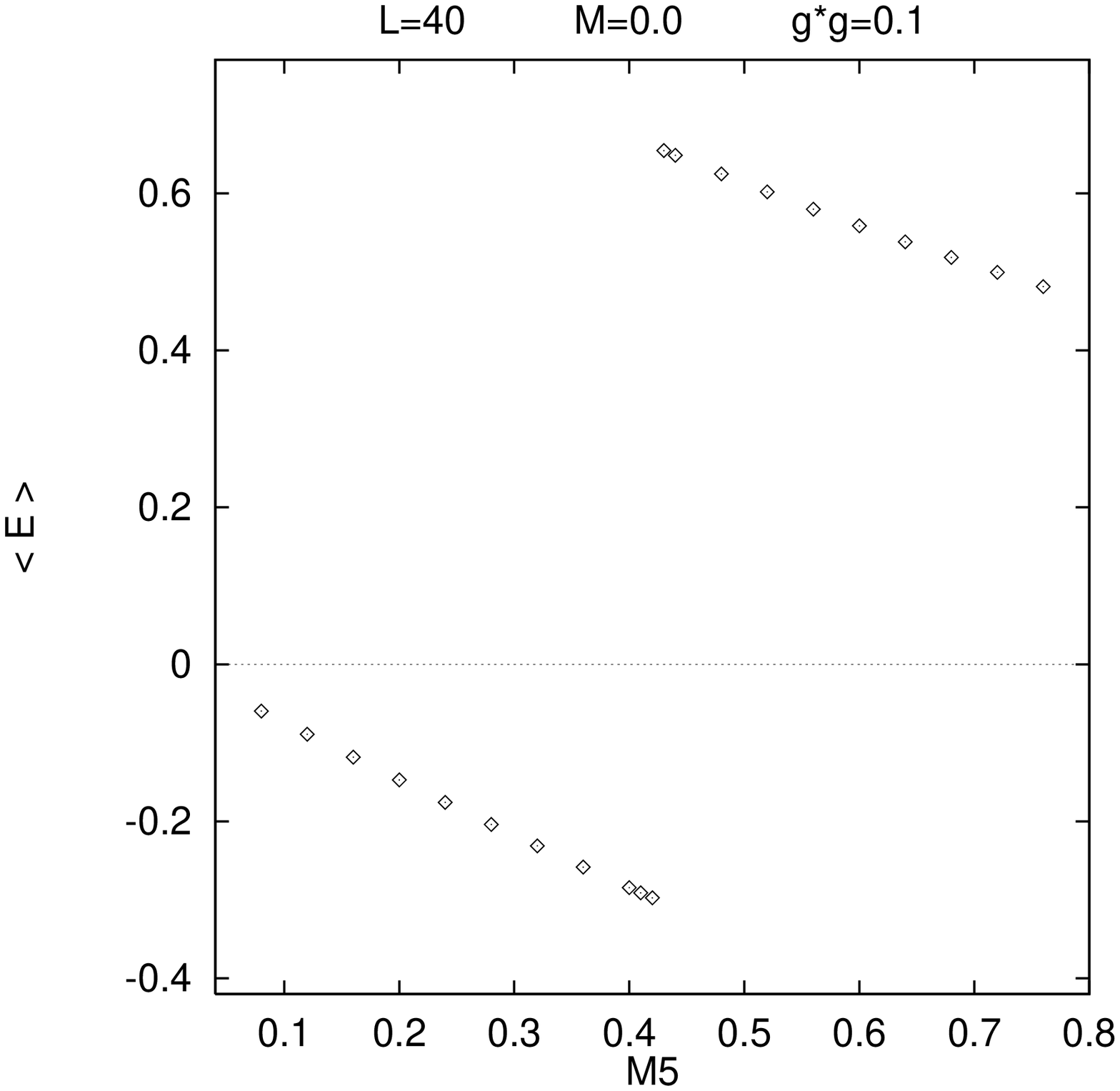}}
\endinsert

\vskip 0.08in
\midinsert
\epsfxsize=0.60\hsize
\epsfysize=0.40\vsize
\centerline{\epsffile {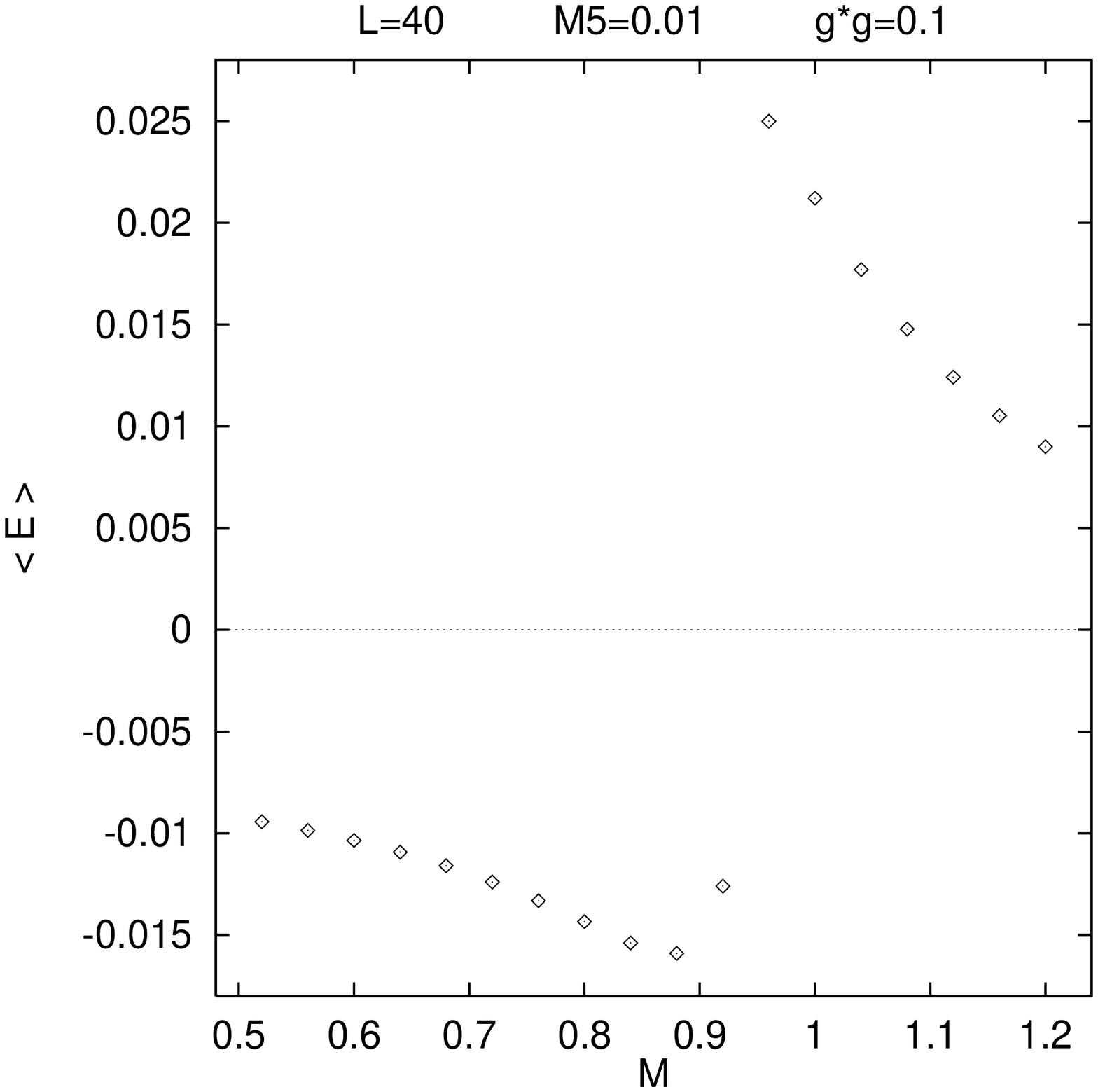}}
\vskip -0.2in
\noindent \narrower {{\bf Fig.9a} (Upper) Electric field vs $M_5$
in the two-flavour model at $g^2=0.1$ and $M=0$ on a lattice with
$40$ sites. {\bf Fig.9b} (Lower) Electric field along the $M$-axis
in the same situation.}
\endinsert

\vfill\eject

\midinsert
\epsfxsize=1.0\hsize
\centerline{\epsffile {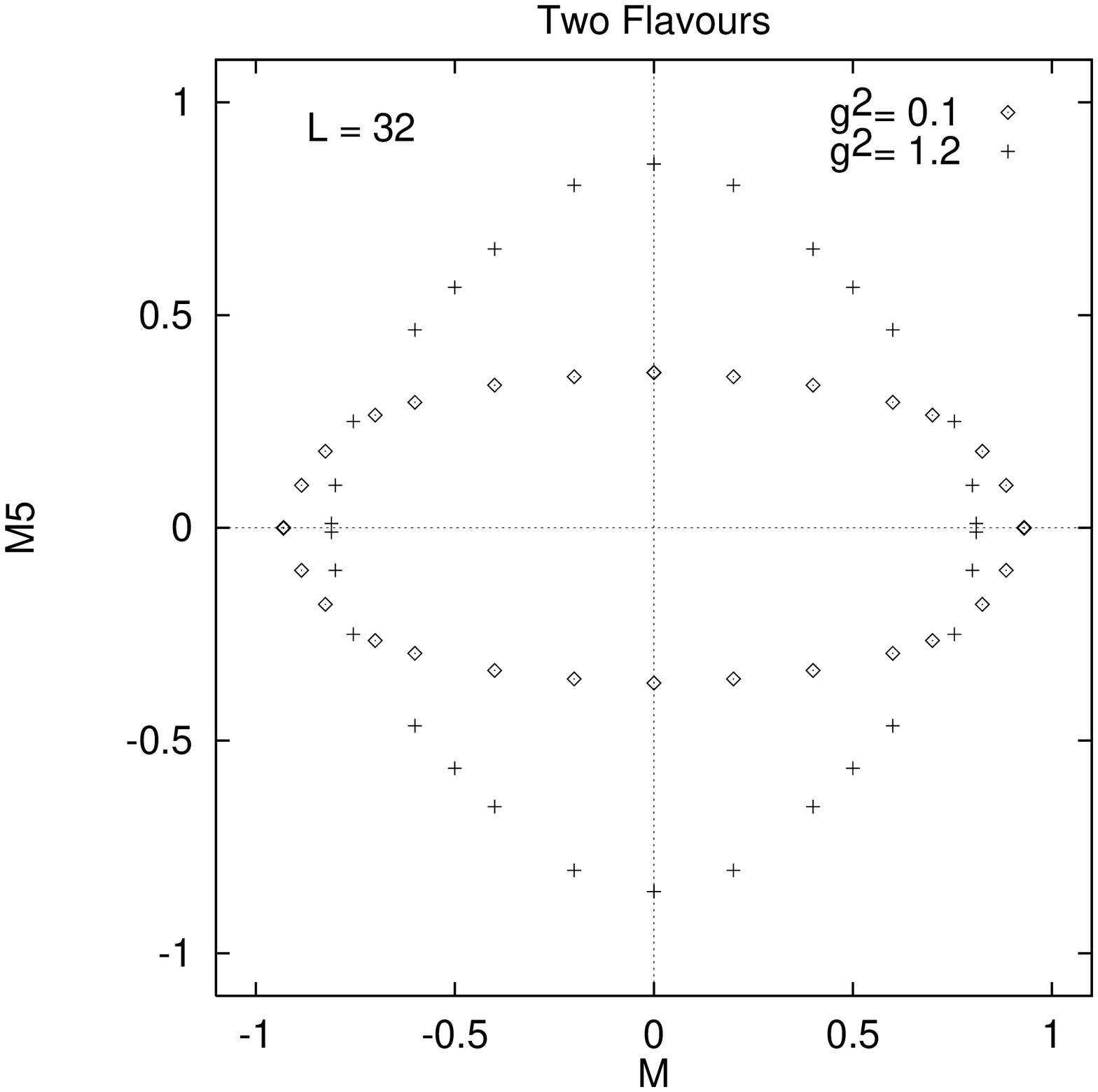}}
\vskip 1.0in
\noindent \narrower {{\bf Fig.10}
Phase diagrams of the two-flavour model in the $M-M_5$ plane
at $g^2=0.1$ and $g^2=1.2$ on a lattice with $32$ sites.
The points represent the positions where electric field in the
Hartree-Fock vacuum changes its sign.}
\endinsert

\vfill\eject

\midinsert
\epsfxsize=0.60\hsize
\epsfysize=0.40\vsize
\centerline{\epsffile {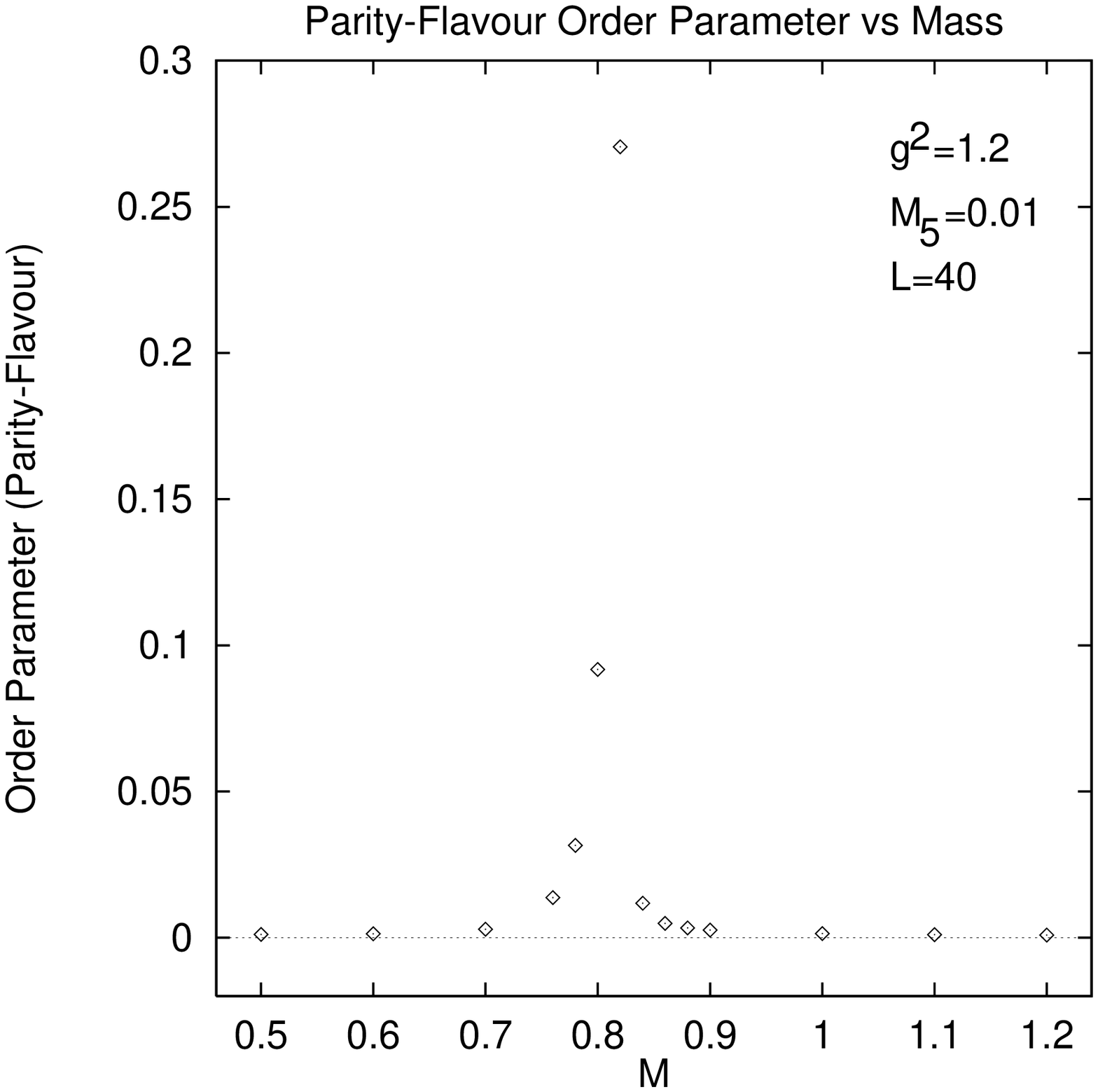}}
\endinsert

\vskip 0.08in
\midinsert
\epsfxsize=0.60\hsize
\epsfysize=0.40\vsize
\centerline{\epsffile {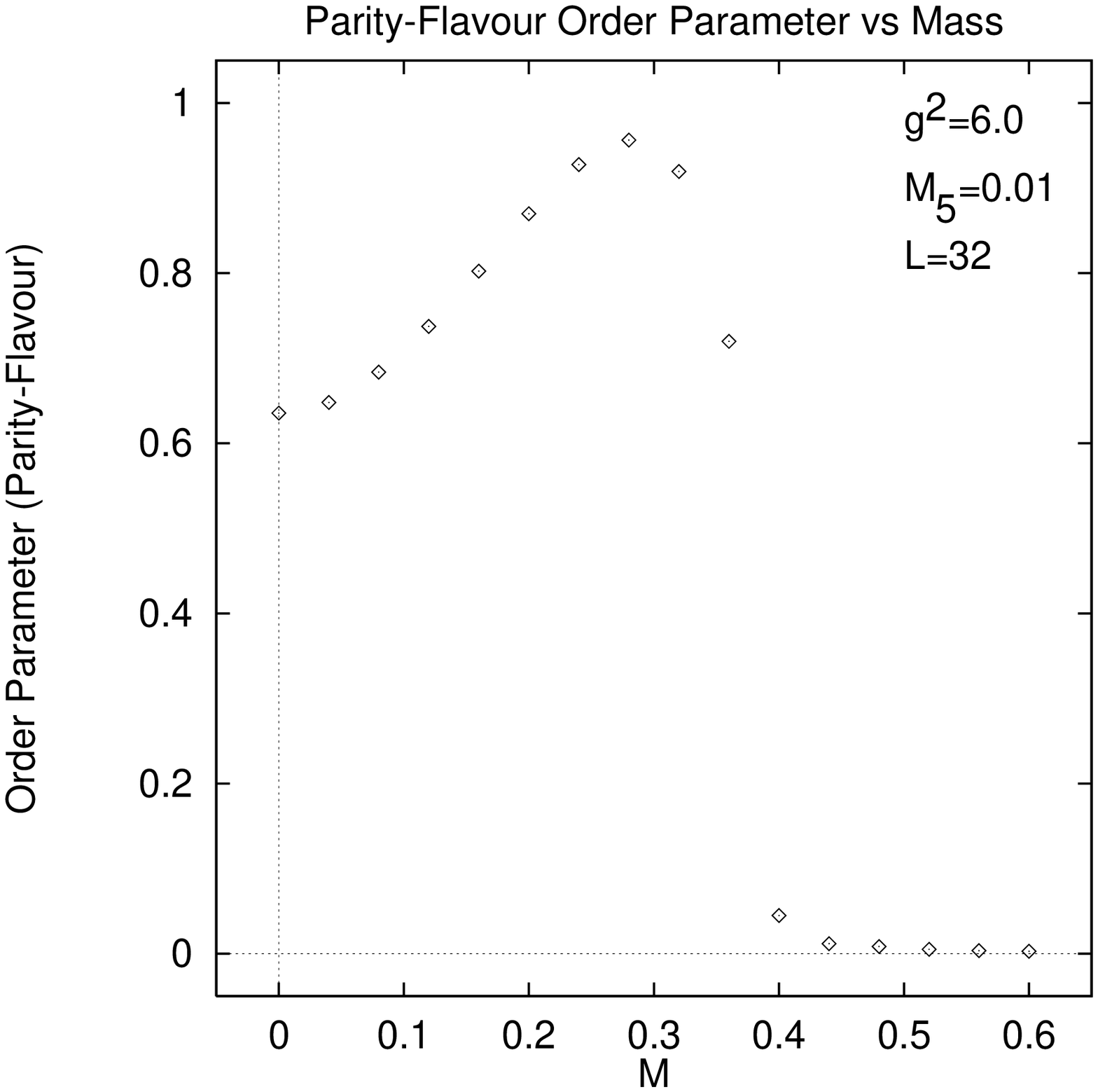}}
\vskip -0.2in
\noindent \narrower {{\bf Fig.11a} (Upper) Expectation value of
$\psibar\gamma_5\tau_3\psi$ at $g^2=1.2$ and $M_5=0.01$ on a lattice with
$40$ sites. {\bf Fig.11b} (Lower) The same expectation at $g^2=6.0$
on a lattice with $32$ sites.}
\endinsert

\vfill\eject

\midinsert
\epsfxsize=1.0\hsize
\centerline{\epsffile {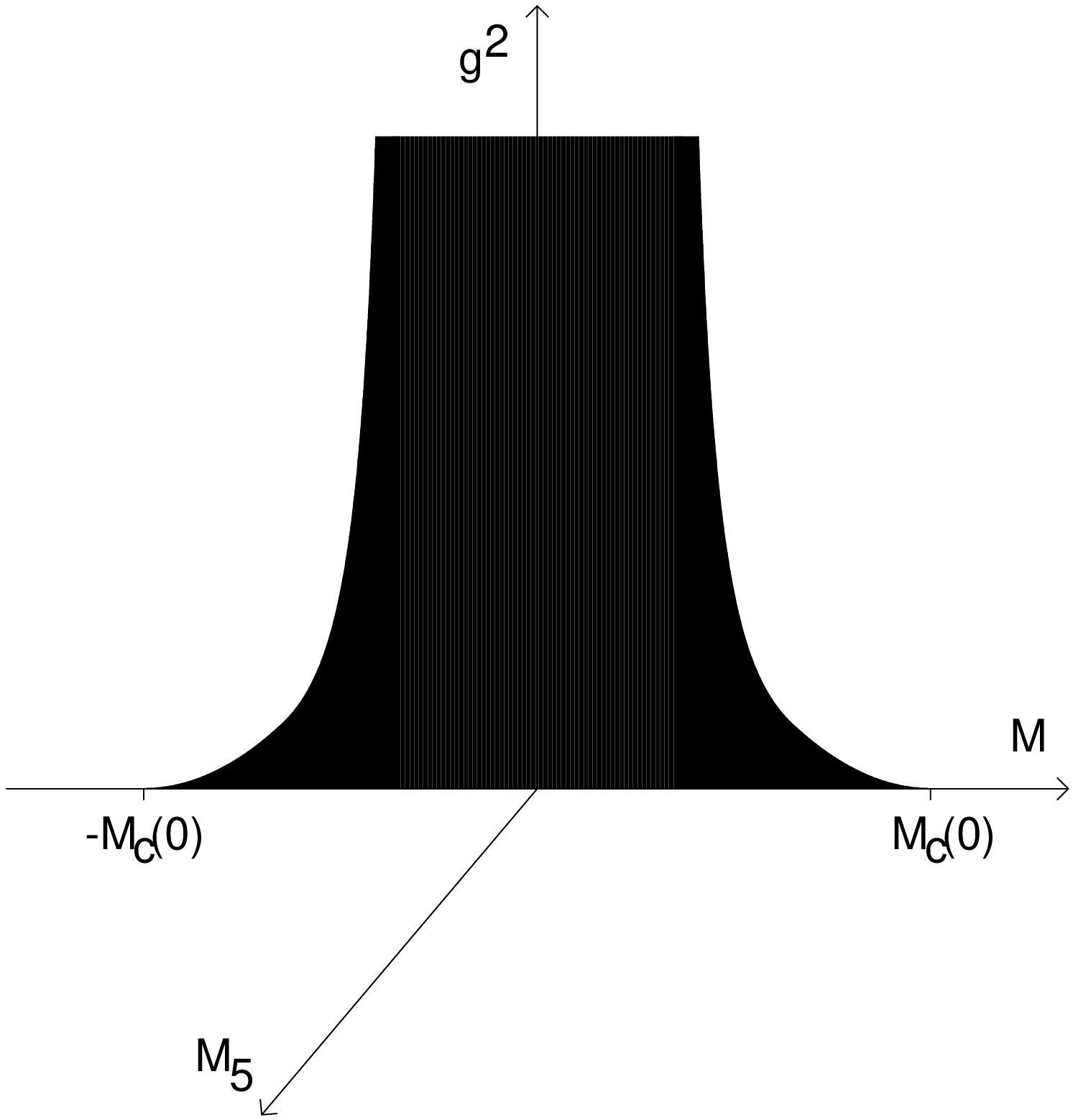}}
\vskip 1.35in
\noindent \narrower {{\bf Fig.12}
The concluded qualitative behaviour of the full H-F phase diagram for
the one-flavour Schwinger model on the lattice with Wilson fermions.
Parity is spontaneously broken in the black sheet, embedded in the
$M-g^2$ plane.}
\endinsert

\vfill\eject

\midinsert
\epsfxsize=0.65\hsize
\epsfysize=0.45\vsize
\centerline{\epsffile {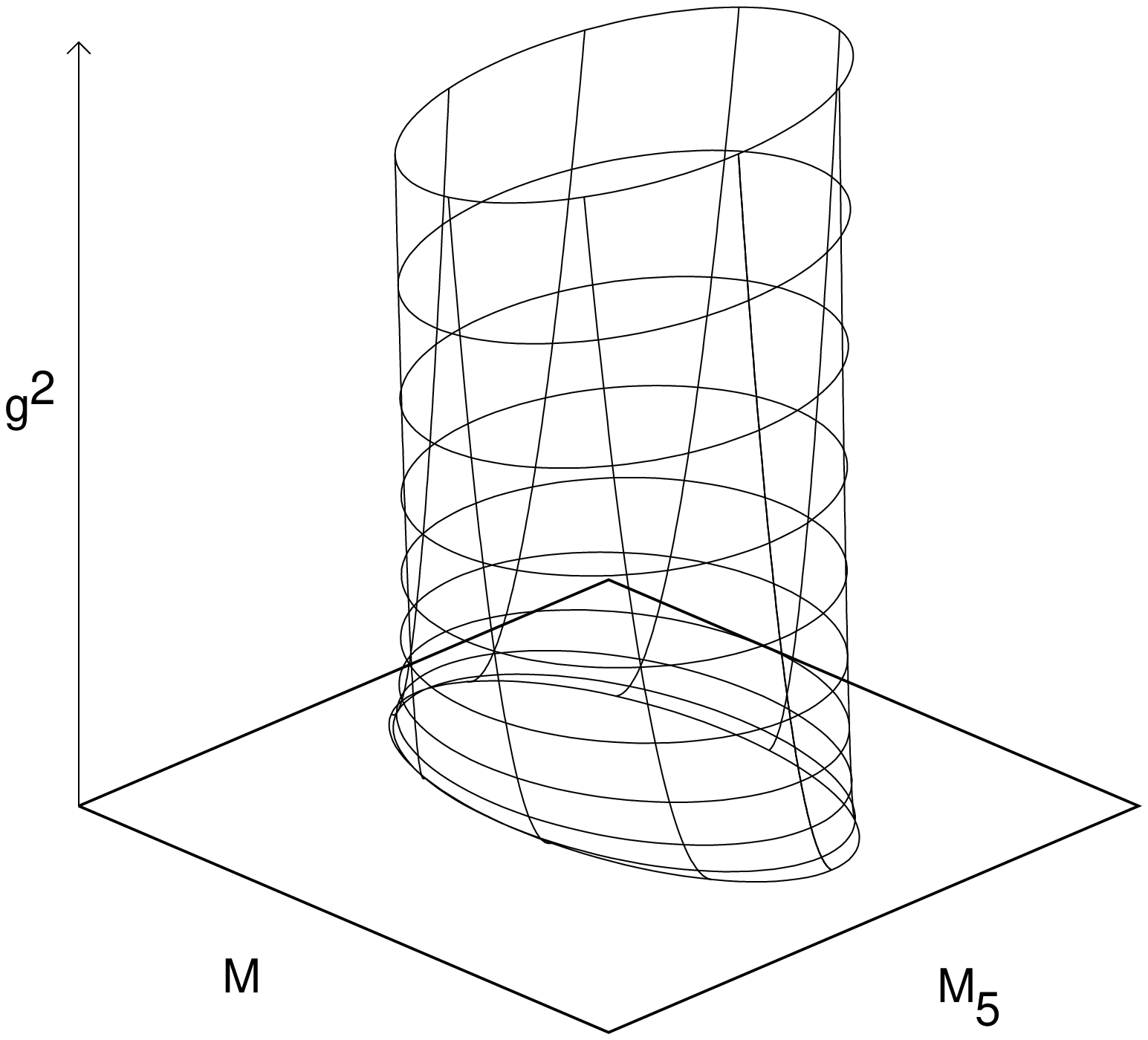}}
\endinsert

\vskip 0.08in
\midinsert
\epsfxsize=0.62\hsize
\epsfysize=0.36\vsize
\centerline{\epsffile {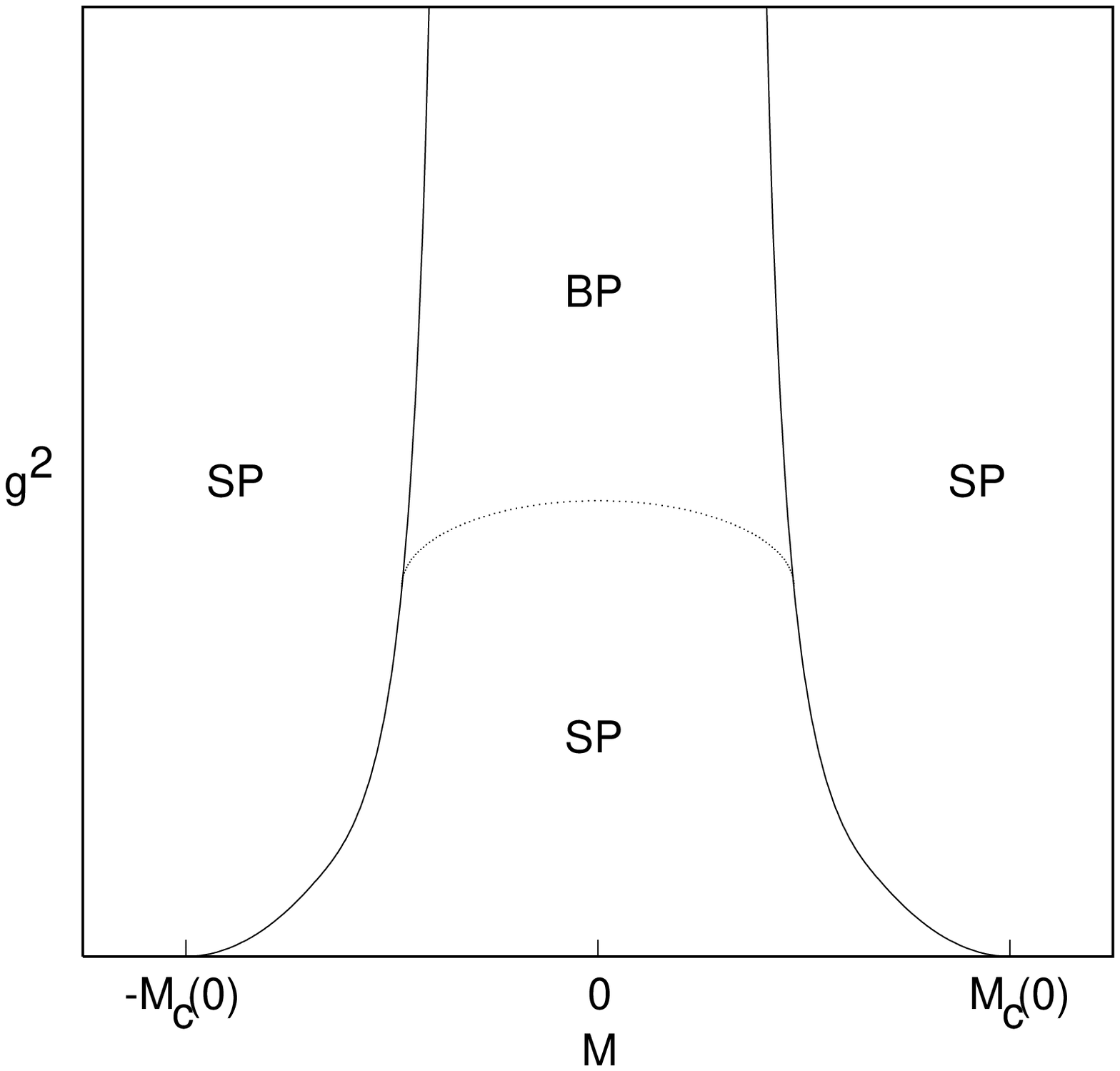}}
\vskip -0.2in
\noindent \narrower {{\bf Fig.13a} (Upper) The concluded qualitative shape
of the surface of ``$\theta =\pi$'' transitions for the model with two
flavours. {\bf Fig.11b} (Lower) The H-F phase diagram of this model
in $M-g^2$ plane. Parity-flavour is spontaneously broken in the
``BP'' region.}
\endinsert

\vfill
\eject
\bye